\documentclass[twocolumn,
trackchanges]{aastex7}
\usepackage{amsmath}
\usepackage{mathrsfs}
\usepackage{mathtools}
\usepackage[retainorgcmds]{IEEEtrantools}
\usepackage{enumitem}
\usepackage{multirow}
\usepackage{fancyhdr,lastpage}

\newcommand{\ud}{\mathrm{d}}

\defcitealias{2020A&A...641A...6P}{\emph{Planck-2018}}

\shorttitle{Multimodality in PTA data: Kination-amplified GWB}
\shortauthors{Li et al.}
\graphicspath{{./}{figures/}}

\begin{document}

\title{Multimodality in the search for new physics in pulsar timing data
and the case of kination-amplified gravitational-wave background from inflation}

\correspondingauthor{Bohua Li}

\author[orcid=0000-0002-3600-0358]{Bohua Li}
\affiliation{Guangxi Key Laboratory for Relativistic Astrophysics, 
School of Physical Science and Technology, Guangxi University,
Nanning, 530004, People’s Republic of China}
\email[show]{bohuali@gxu.edu.cn}

\author[orcid=0000-0001-8510-2812]{Joel Meyers}
\affiliation{Department of Physics, Southern Methodist University, Dallas, 
TX 75275, USA}
\email{jrmeyers@mail.smu.edu}

\author[orcid=0000-0002-0410-3045]{Paul R. Shapiro}
\affiliation{Department of Astronomy, The University of Texas at Austin, Austin, 
TX 78712, USA}
\email{shapiro@astro.as.utexas.edu}

\begin{abstract}

We investigate the kination-amplified inflationary gravitational-wave background (GWB)
interpretation of the signal recently reported
by various pulsar timing array (PTA) experiments.
Kination is a post-inflationary phase in the expansion history
dominated by the kinetic energy of some scalar field, 
characterized by a stiff equation of state $w=1$.
Within the inflationary GWB model, we identify two modes
which can fit the current data sets (NANOGrav and EPTA) with equal likelihood:
the kination-amplification (KA) mode
and the ordinary, non-kination-amplification (no-KA) mode.
The multimodality of the likelihood motivates a Bayesian analysis with nested sampling.
We analyze the free spectra of current PTA data
and mock free spectra constructed with higher signal-to-noise ratios, 
using nested sampling.
The analysis of the mock spectrum designed to be consistent
with the best fit to the NANOGrav 15 yr (NG15) data
successfully reveals the expected bimodal posterior for the first time
while excluding the reheating mode
that appears in the fit to the current NG15 data, 
making a case for our correct and comprehensive treatment
of potential multimodal posteriors arising from future PTA data sets.
The resultant Bayes factor is $\mathcal{B}\equiv\mathcal{Z}_\mathrm{no-KA}/\mathcal{Z}_\mathrm{KA}=2.9\pm1.9$,
indicating comparable statistical significance between the two modes.
Given the theoretical model-building challenges
of producing highly blue-tilted primordial tensor spectra,
the KA mode has the advantage of requiring less blue primordial spectra,
compared with the no-KA mode.
The synergy between future cosmic microwave background polarization, 
pulsar timing and laser interferometer measurements of gravitational waves will
help resolve the ambiguity implied by the multimodal posterior in PTA-only searches.

\end{abstract}



\section{Introduction} \label{sec:intro}

The field of gravitational-wave (GW) astronomy recently witnessed a major breakthrough
brought by various pulsar timing array (PTA) collaborations, 
NANOGrav \citep{2023ApJ...951L...8A}, EPTA \& InPTA \citep{2023A&A...678A..50E}, 
PPTA \citep{2023ApJ...951L...6R} and CPTA \citep{2023RAA....23g5024X}, 
who concurrently reported strong evidence
for a stochastic gravitational-wave background (SGWB) signal
in the nanoHertz frequency band.
The signature interpulsar Hellings-Downs (HD) correlation curve \citep{1983ApJ...265L..39H}
has been measured up to 4.6$\sigma$ statistical significance.

While the confirmation of this signal is underway,
interpretation of the existing data has already drawn tremendous attention
among the broader astrophysics and high-energy physics community.
Beyond the standard interpretation that the nanohertz SGWB is sourced
by inspiraling supermassive black hole binaries
\citep[SMBHBs;][]{1995ApJ...446..543R, 2003ApJ...583..616J, 2004ApJ...611..623S},
new-physics models may provide a better fit to the measured free spectrum
\citep{2023ApJ...951L..11A, 2024A&A...685A..94E, 2023arXiv231206756S}.
Hence, this potential SGWB signal offers abundant opportunities 
for early-Universe cosmology
\citep[e.g.,][]{2023JHEAp..39...81V, 2024PhRvL.132q1002F, 2024PhRvD.109b3522E}. 

Cosmological interpretations of the PTA data can be roughly
classified as inflationary or non-inflationary signals.
Inflationary SGWBs are expected from tensor vacuum fluctuations of the inflaton
\citep{1974ZhETF..67..825G, 1979JETPL..30..682S, 1984NuPhB.244..541A}
or spectator fields \citep{2012PhRvD..85b3534C, 2016JCAP...01..041N, 2018PhRvD..97b3532C},
and their post-inflationary amplification
\citep{1998PhRvD..58h3504G, 2008PhLB..668...44G, 2008PhRvD..78d3531B, 2017PhRvD..96f3505L}.
Non-inflationary SGWBs are produced by causal physics
such as first-order phase transitions \citep{1994PhRvD..49.2837K, 2014PhRvL.112d1301H, 2015PhRvL.115r1101S, 2021NatAs...5.1268M, 2021PhRvL.127y1302A, 2021PhRvL.127y1303X}, topological defects \citep{2018PhRvD..97l3505C, 2020JCAP...04..034A, 2021PhRvL.126d1304E, 2021PhRvL.126d1305B, 2021PhRvD.103f3031R}, 
preheating \citep{1997PhRvD..56..653K, 2007PhRvL..99v1301E, 2010JCAP...04..021J},
scalar fluctuations \citep{2007PhRvD..75l3518A, 2007PhRvD..76h4019B, 2019PhRvL.122t1101C, 2021Univ....7..398D, 2023SciBu..68.2929Y, 2023PhRvL.131t1401F, 2023JCAP...11..071L, 2024PhRvD.109f1301L}, etc.

In view of the plenitude of models to be tested by the same PTA data sets, 
carefully implemented methodology for model selection/comparison 
is key to drawing robust conclusions about any potential new physics.
Current model selection methods are mainly based on the Bayes factor, 
$\mathcal{B}_{10}\equiv\mathcal{Z}_1/\mathcal{Z}_0$,
defined as the ratio of the evidence of two models.
However, it is understood that the current PTA data are not informative enough 
to make the Bayes factors free from the ``prior volume effect'',
i.e., their exact values depend strongly on 
the range of priors adopted for the parameters
\citep{2023ApJ...951L..11A, 2024A&A...685A..94E}.
Better measurements of the SGWB spectrum are required, therefore, 
to minimize the prior volume effect
and derive stable Bayes factors for new-physics models.

In this paper, we shall address
another subtlety in the analysis and interpretation of PTA data
regarding the search for new physics, 
namely the \emph{multimodal} posterior distribution of model parameters.
Such a posterior contains disjoint high-likelihood clusters (modes),
while each mode may bear a distinctive physical interpretation.
As an example, \citet{2023ApJ...951L..11A} performed a Bayesian fit
to the NANOGrav 15yr (NG15) data using the inflationary SGWB model
that allows for a primordial blue tilt
(positive tensor spectral index $n_\mathrm{t}$),
and found a bimodal posterior where the signal can be explained by
either the tensor perturbations that reentered the horizon
during radiation domination (RD)
or those that reentered during reheating.
However, this ``reheating mode'' has a spectral index of $n_\mathrm{t}-2$
in the SGWB spectrum today, which requires a notably large $n_\mathrm{t}$
to fit the PTA signal. Its best-fit parameters may thus be in conflict
with the linearity of primordial tensor fluctuations, as we show in this paper.
Instead, we herein consider another case for a potential multimodal posterior:
the inflationary SGWB with kination amplification
\citep{2011PhRvD..84l3513K, 2017PhRvD..96f3505L, 2018CQGra..35p3001C, 2019JCAP...08..011F, 2021arXiv211101150G, 2022JHEP...09..116C}.
Kination is an early phase in the expansion history
in which the Universe is dominated by the kinetic energy of some scalar field 
\citep{1993PhLB..315...40S, 1997PhRvD..55.1875J, 2020PhRvL.124y1802C}, 
also dubbed a ``stiff phase,'' since the equation of state (EoS) during kination
is that of a relativistically-stiff perfect fluid,
i.e. $w\equiv\bar P/\bar\rho=1$ \citep{2014PhRvD..89h3536L}.
When kination is present, the kination-amplified 
(also referred to as ``stiff-amplified'') part of the inflationary SGWB spectrum today
has index $n_\mathrm{t}+1$, instead of the primordial index $n_\mathrm{t}$, 
which introduces a 
new mode
that can fit the same PTA data
within the inflationary SGWB model \citep{2021JCAP...10..024L}.

In contrast to the prior volume effect, 
multimodality in the posterior becomes more substantial as data get more informative.
Therefore, it is crucial to take into account the possibility of multimodality
in the analysis of future PTA data.
In addition to the common model selection exercise, 
``mode selection'' within one model needs to be performed when possible.
To this end, nested sampling \citep{2006Skilling} provides
a robust method to explore multimodal posteriors.
We here apply the nested sampler
implemented by the \verb|PolyChord| package 
\citep{2015MNRAS.450L..61H, 2015MNRAS.453.4384H} and evaluate the Bayes factor
between the two modes of the inflationary SGWB
(with kination amplification or without).


The purpose of this paper is to demonstrate the necessity of
considering multimodality (and hence nested samplers)
when interpreting the current and future PTA data sets,
by the case study of the kination-amplified inflationary SGWB model,
a noteworthy model per se.
The rest of the paper is organized as follows.
We describe our model in Section~\ref{sec:kination}
and the data used in this paper in Section~\ref{sec:data}. 
Our results on the bimodal posteriors are shown in Section~\ref{sec:results}
and relevant discussions are presented in Section~\ref{sec:discussion}.
We conclude in Section~\ref{sec:conclusion}.

\section{Model: Kination-amplified SGWB}\label{sec:kination}

\begin{figure*}[ht!]
\resizebox{\textwidth}{!}{\includegraphics{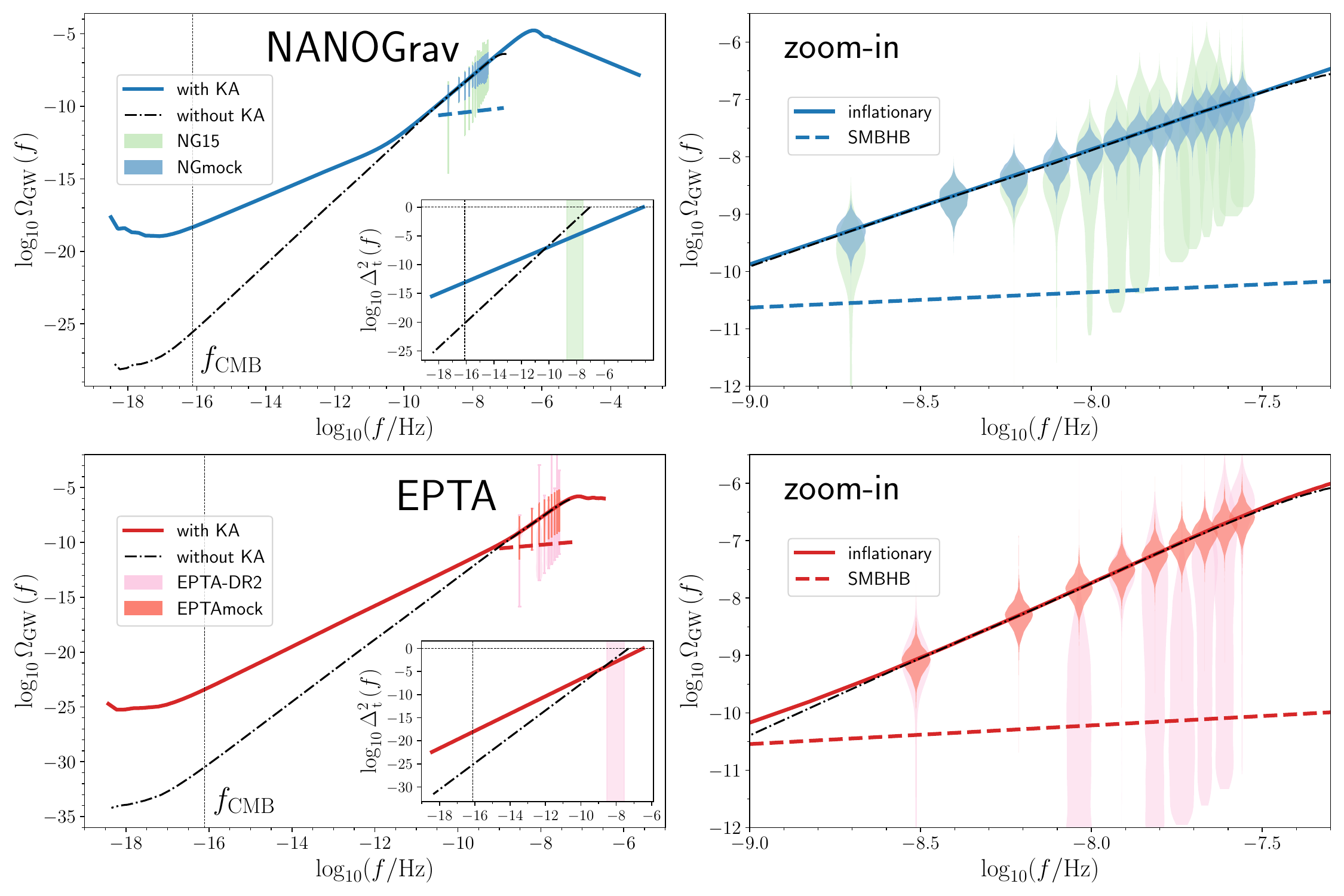}}
\caption{Inflationary SGWB spectra with or without kination amplification.
The free spectra obtained from the current PTA data sets 
as well as our mock free spectra are indicated by the shaded violins:
the NG15 and NGmock spectra in the upper panels,
the EPTA-DR2 and EPTAmock spectra in the lower panels.
The right panels zoom in on the free spectra.
In all panels, the inflationary SGWB spectrum curves 
illustrate the best fits to the corresponding \emph{mock spectra},
colored solid curves for the kination-amplification (KA) mode 
and black dash-dotted curves for the no-KA mode.
The colored dashed lines indicate the accompanying SMBHB component in the KA mode.
The insets in the left panels show the blue-tilted primordial tensor spectra,
where the nonlinearity cutoffs are illustrated (cf. Appendix~\ref{app:model}).
The vertical shaded regions denote the PTA band.
}
\label{fig:spectra}
\end{figure*}

The description of the kination-amplified inflationary SGWB
follows closely \citet{2021JCAP...10..024L}.
Tensor metric perturbations produced by inflation
are assumed to satisfy a power-law initial power spectrum, 
$\Delta_\mathrm{t}^2(f)=A_\mathrm{t}\,(f/f_\mathrm{CMB})^{n_\mathrm{t}}$, 
where the tensor amplitude $A_\mathrm{t}$ is related to the scalar amplitude
by the tensor-to-scalar ratio, $r\equiv A_\mathrm{t}/A_\mathrm{s}$, 
and $k_\mathrm{CMB}\equiv 2\pi f_\mathrm{CMB}/c=0.05\,\text{Mpc}^{-1}$ is the cosmic microwave background (CMB) 
pivot scale \citep{2020A&A...641A..10P}. 
Canonical single-field slow-roll inflation
generates a slightly red-tilted primordial tensor spectrum
that satisfies the consistency relation, $n_\mathrm{t}=-r/8$.
However, various nonstandard scenarios can predict a significant blue tilt
\citep{2004PhRvD..69l7302B, 2008JCAP...08..012S, 2014PhRvL.113w1301C, 2015NuPhB.900..517C, 2019PhLB..789..215F, 2022JCAP...07..034B}.
In this work, we relax the consistency relation
and treat $n_\mathrm{t}$ as an independent parameter.

After reentering the horizon, inflationary tensor perturbations form an SGWB. 
Its energy spectrum today is given by
\begin{equation}\label{eq:Omegagw}
    \Omega_\mathrm{GW}(f) \equiv \frac{\ud\Omega_\mathrm{GW}}{\ud\ln f}
    = \Delta_\mathrm{t}^2(f) \frac{(2\pi f)^2\,T^2(f)}{12H_0^2},
\end{equation}
where $T(f)\equiv h(f)/h_\mathrm{ini}(f)$ is the tensor transfer function,
$h(f)$ is the Fourier-space tensor perturbation today
and $h_\mathrm{ini}(f)$ is its initial superhorizon value.
In general, the tensor transfer function is nontrivial
and its accurate theoretical prediction must be calculated numerically.
We solve the exact tensor wave equation for each frequency
iteratively to take account of
the backreaction of the SGWB energy density on the expansion history
(cf. Appendix~\ref{app:model}).
Our numerical scheme is described in \citet{2021JCAP...10..024L}
and now available as the \verb|stiffGWpy|
package (link provided at the end of the paper).
On the other hand, the tensor transfer function follows a simple power law
for wavelengths that reentered the horizon
during an era of a \emph{constant} EoS,
so that the GW spectrum in the corresponding frequency range can be expressed as
\begin{equation}\label{eq:Omegagw_pl}
    \Omega_\mathrm{GW}(f) \propto 
    \Delta_\mathrm{t}^2(f)\left(\frac{f\,a_f}{H_0}\right)^2
    \propto f^{\,n_\mathrm{t}+\frac{2(3w_f-1)}{1+3w_f}},
\end{equation}
where $a_f$ is the scale factor at which the tensor perturbation
of frequency $f$ reentered the horizon, $2\pi f\equiv a_f H(a_f)$,
and $w_f$ is the EoS parameter then \citep{2008PhRvD..77f3504B, 2017PhRvD..96f3505L}.
Therefore, on top of the primordial tensor spectral index,
kination/stiff amplification introduces
an additional blue tilt for frequencies with $w_f=1$, 
so that $\Omega_\mathrm{GW}(f)\propto f^{\,n_\mathrm{t}+1}$ in this frequency range,
as shown by Eq.~(\ref{eq:Omegagw_pl}).

We consider a simple extension to the base $\Lambda$CDM model, 
in which the thermal history of the Universe starts in kination and then
transitions into RD prior to Big Bang nucleosynthesis (BBN), 
$w_\mathrm{RD}=1/3$.
Kination is modeled by a stiff fluid ($w_\mathrm{s}=1$) 
and the kination-to-radiation transition is parameterized by 
$\kappa_{10}\equiv (\rho_\mathrm{s}/\rho_\gamma) |_{T=10~\mathrm{MeV}}$, 
the ratio of the stiff-fluid density to the photon density at $10$~MeV
\citep{2010PhRvD..82h3501D}.
We assume that the inflationary phase ends into a prolonged reheating epoch 
dominated by the coherent oscillations of the inflaton field, so $w_\mathrm{re}=0$.
The end of reheating marks the onset of kination.
In this work, we fix all base-$\Lambda$CDM parameters 
to their \citetalias{2020A&A...641A...6P} best-fit values,
so our model contains the following free parameters only:
$\{r, n_\mathrm{t}, \kappa_{10}, T_\mathrm{re}, \Delta N_\mathrm{re}\}$, 
where $T_\mathrm{re}$ is the temperature at the end of reheating 
and $\Delta N_\mathrm{re}$ is the number of $e$-folds during reheating.
The primordial tensor spectrum in this model is given by
the power law described above \emph{only} for $f<f_\mathrm{cut}$,
where $f_\mathrm{cut}$ is a UV cutoff on the primordial power spectrum
which excludes the pathological nonlinear regime at high frequencies;
see the insets in the left panels of Fig.~\ref{fig:spectra}.
Note that this new UV cutoff is necessary for a consistent description
of a blue-tilted primordial spectrum,
but has unfortunately been neglected in some previous studies.
Further details of the model are elaborated in Appendix~\ref{app:model}.


Altogether, our model results in a \emph{broken-power-law} energy spectrum
for the kination-amplified inflationary SGWB today, $\Omega_\mathrm{GW}(f)$
\citep{2021JCAP...10..024L, 2022PhRvD.105d3520B},
as illustrated in Fig.~\ref{fig:spectra}.
Wavelengths that reentered during RD, kination and reheating have spectral indices of
$n_\mathrm{t}$, $n_\mathrm{t}+1$ and $n_\mathrm{t}-2$, respectively.
If $\kappa_{10}\gtrsim 10^{-2}$,
the kination phase is important for the PTA frequency band.
If $\kappa_{10}\ll10^{-2}$, the frequency corresponding to
the kination-to-radiation transition exceeds the PTA band,
so that our inflationary SGWB model effectively reduces to the ordinary, 
non-kination-amplified mode in the PTA band.

\section{Data analysis}\label{sec:data}

The presence of an SGWB induces an HD-correlated, common-spectrum (red-noise) process
to the timing residuals/delays measured by PTAs. 
The GW spectrum today is thus related to the timing-residual power spectral density
\citep[e.g.,][]{2021NatAs...5.1268M,2023ApJ...951L...8A}, such that
\begin{equation}
    \Omega_\mathrm{GW}(f) = \frac{2\pi^2 f^2}{3H_0^2}h_c^{\,2}(f)
    =\frac{8\pi^4 f^5}{H_0^2}T_\mathrm{obs}\,\Phi(f),
\end{equation}
where $h_c(f)$ is the characteristic strain of the SGWB,
$T_\mathrm{obs}$ is the timing baseline
and $\Phi(f)$ is the variance of the Fourier components of timing residuals.

In this paper, we analyze the free spectra obtained from the NG15 \citep{2023NG15_KDE}
and the EPTA DR2new \citep[EPTA-DR2;][]{2023EPTA_DR2} data sets, 
illustrated by the light-shaded violins in Fig.~\ref{fig:spectra}.
For each free spectrum, 
we perform a Bayesian search for the kination-amplified inflationary SGWB
in the presence of a GW foreground from SMBHBs. 
We do not multiply the NG15 and EPTA-DR2 likelihoods in our analysis
but rather consider them separately,
since the two data sets contain overlapping pulsars \citep{2024ApJ...966..105A}.
A proper combination of various PTA data sets
will be provided by the forthcoming IPTA’s third data release.

Beyond current data sets, we consider mock data
with improved signal-to-noise ratios (S/Ns)
to investigate the sensitivity required for future PTAs
to overcome the prior volume effect and detect the kination-amplification mode.
We generate two power-law mock free spectra with high S/Ns, 
labeled NGmock and EPTAmock,
each with the mean spectral index and frequency bins
of the NG15 and EPTA-DR2 data, respectively.
They are illustrated by the uniform-shaped, dark-shaded violins
in Fig.~\ref{fig:spectra}.
While the sensitivity of realistic PTAs typically has 
a non-uniform frequency dependence, 
each of our mock spectra assumes a uniform uncertainty
across its respective frequency bins, set to the smallest uncertainty
of the currently measured free spectrum.\footnote{
In either the NG15 and EPTA-DR2 free spectrum, it is the second lowest frequency bin
that currently has the highest S/N (see Fig.~\ref{fig:spectra}).}
Fig.~\ref{fig:spectra} shows that both the kination-amplification (KA) mode
and the non-kination-amplification (no-KA) mode
can fit the mock spectra equally well.
Therefore, we expect the emergence of bimodality
in the corresponding analysis results.
Note that here the no-KA mode
exclusively refers to the case in which the PTA signal
is fit by the wavelengths that reentered the horizon during RD,
since the reheating mode of spectral index $n_\mathrm{t}-2$
cannot simultaneously fit the mock spectra and satisfy the linearity threshold,
as the insets in the left panels of Fig.~\ref{fig:spectra} imply.

Our Bayesian analyses are performed through nested sampling
as well as the traditional Markov Chain Monte Carlo (MCMC) sampling, 
using the \verb|cobaya| code
\citep{2019ascl.soft10019T, 2021JCAP...05..057T}. 
We assume uniform prior distributions for the aforementioned parameters
concerning the inflationary SGWB and kination:
$n_\mathrm{t}\in[-1,\,6]$, $\log_{10}\kappa_{10}\in[-7,\,3]$,
$\log_{10}(T_\mathrm{re}/\mathrm{GeV})\in[-3,\,7]$,
$\Delta N_\mathrm{re}\in[0,\,40]$.
For the tensor-to-scalar ratio,
we adopt $\log_{10}r\in[-40,\,0]$ or $[-20,\,0]$ for the NG15 \& NGmock analyses,
and $\log_{10}r\in[-25,\,0]$ for the EPTA-DR2 \& EPTAmock analyses.
By contrast, a multivariate Gaussian prior is used
for parameters concerning the SGWB from SMBHBs (cf. Appendix~\ref{app:model}),
adopting the same model as in \citet{2023ApJ...951L..11A},
$h_c(f)=A_\mathrm{BHB}\,(f/f_\mathrm{yr})^{(3-\gamma_\mathrm{BHB})/2}$.

\begin{figure}[ht!]
\gridline{\fig{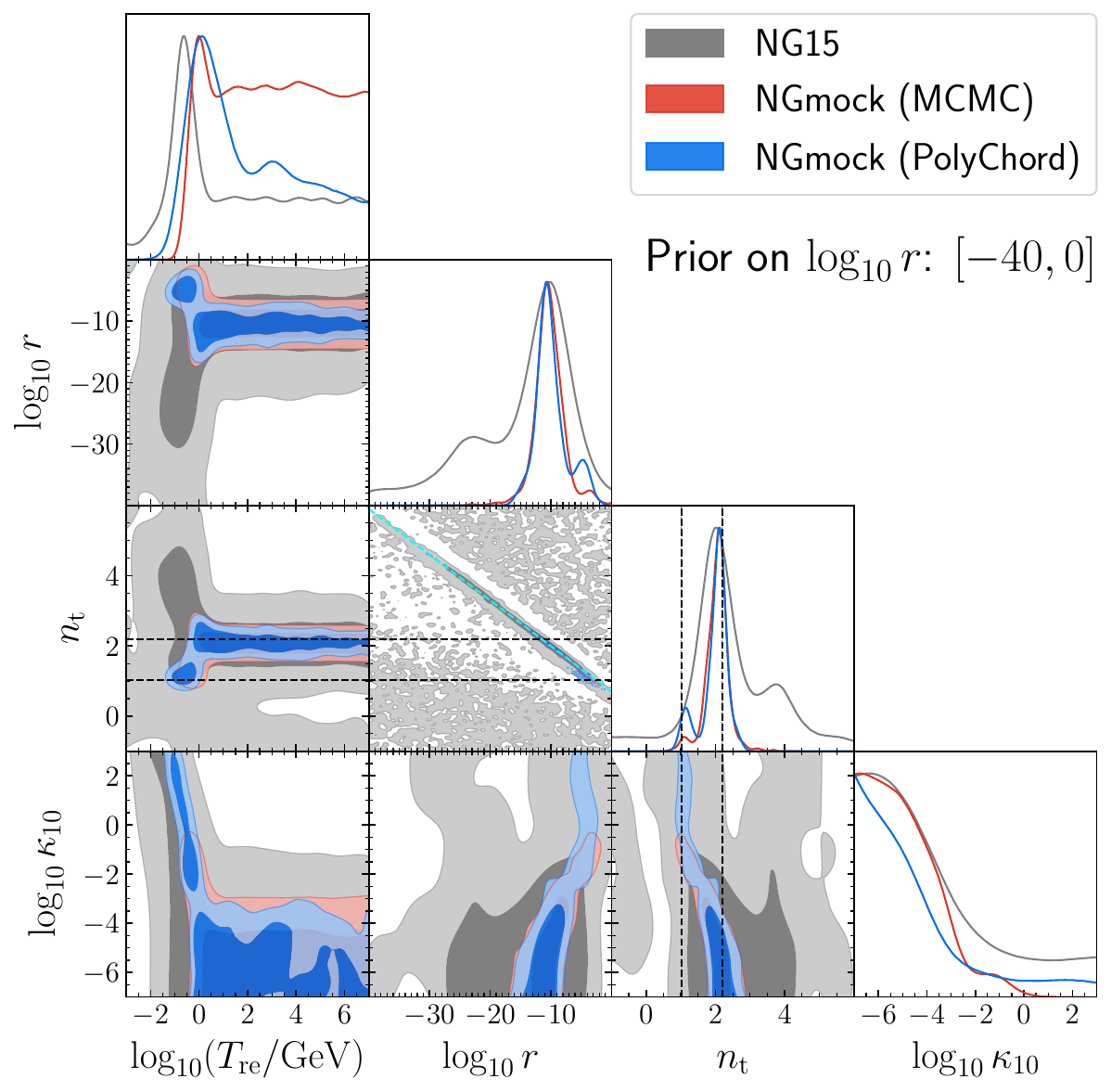}{0.95\columnwidth}{}}
\gridline{\fig{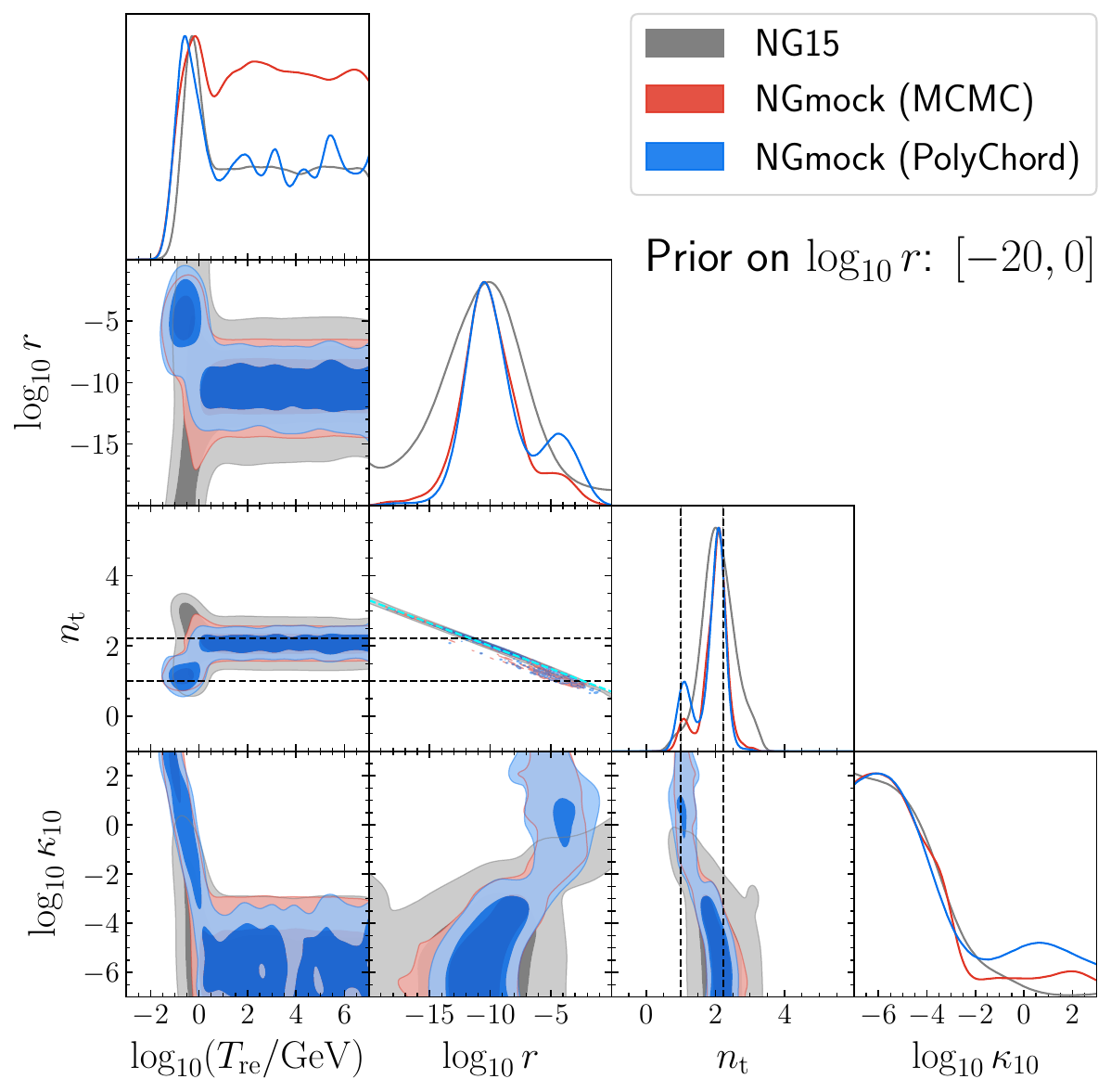}{0.95\columnwidth}{}}
\caption{Constraints on the inflationary SGWB model
with possible kination amplification from the NG15 and NGmock data.
The upper and lower figures show the result
obtained using a wide prior on $\log_{10}r$
and that obtained using a narrow prior, respectively.
Only a subset of the model parameters is presented, whose posteriors are illustrated 
by the 68\% and 95\% 2D contours in the off-diagonal panels 
and the 1D marginalized distributions in the diagonal panels.
The black dashed lines indicate the best-fit values of $n_\mathrm{t}$
to the mock spectra,
for both the KA and the no-KA modes (cf. Table~\ref{tab:mock}).
The cyan dashed lines in the $n_\mathrm{t}-\log_{10}r$ panels 
denote the linear fits to the 2D contours; see Eq.~(\ref{eq:lineprimordial}).
\label{fig:triangle}}
\end{figure}

\section{Results: bimodal posteriors}\label{sec:results}

\begin{figure}[ht!]
\epsscale{1.1}
\plotone{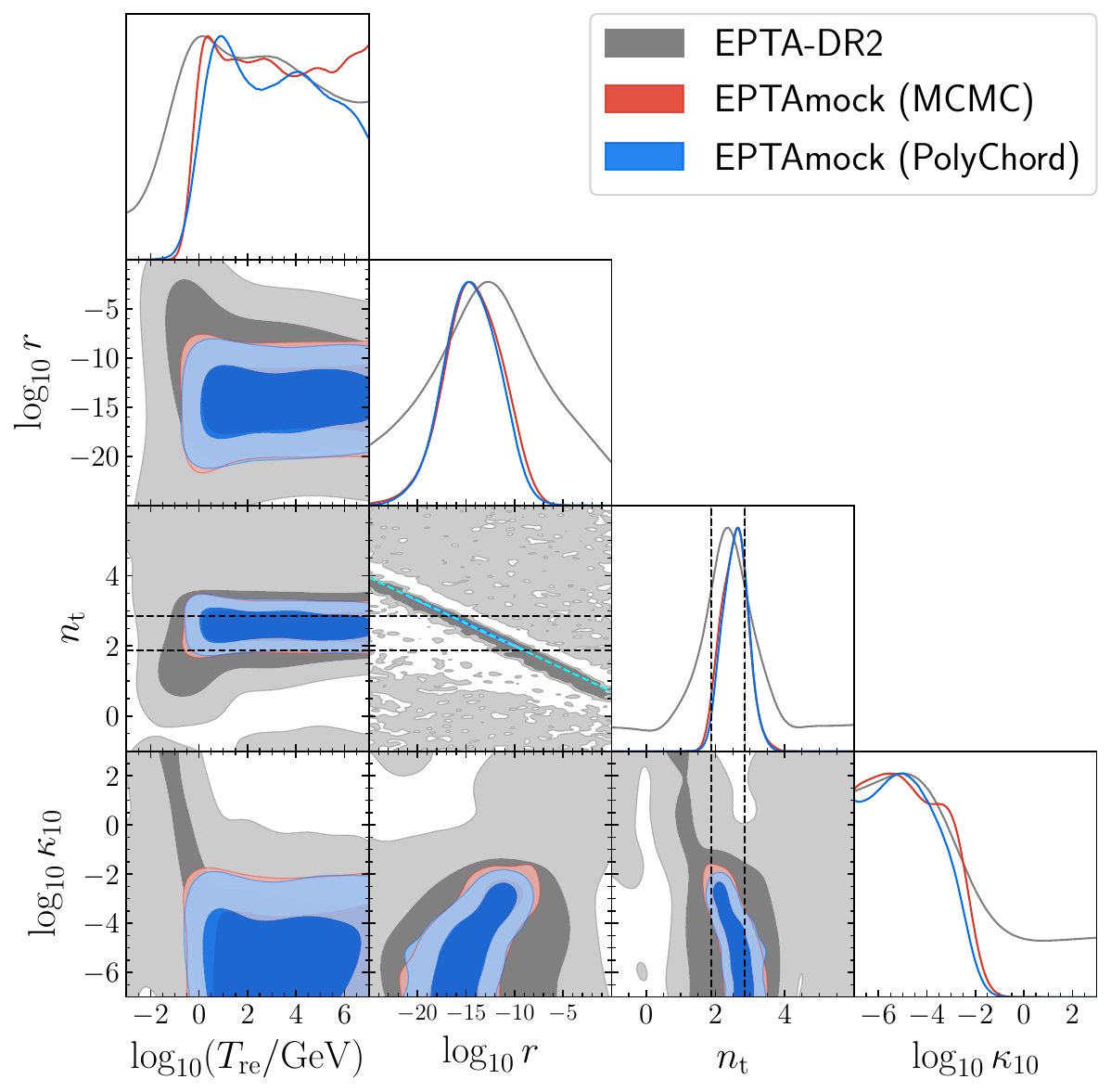}
\caption{Same as in Fig.~\ref{fig:triangle}, 
but for the EPTA-DR2 and EPTAmock data.
}
\label{fig:triangle_EPTA}
\end{figure}

We now present the results of our Bayesian inference.
Fitting the current NG15 and EPTA-DR2 spectra to our model
essentially confirms the results in \citet{2023ApJ...951L..11A}
and \citet{2024A&A...685A..94E}, respectively.
In these cases, the posterior distributions of the parameters are indicated by
the gray contours in Figs.~\ref{fig:triangle} \& \ref{fig:triangle_EPTA}.
While some of the 2D posteriors hint at the presence of the KA mode
(e.g., the upper left region within the $\log_{10}\kappa_{10}-n_\mathrm{t}$ contour
with $\kappa_{10}\gtrsim 10^{-2}$ and $n_\mathrm{t}\sim1$),
all the 1D marginalized distributions of the model parameters
exhibit dominance of the ordinary, no-KA mode.
In this regime, the PTA data are basically fit
by an inflationary SGWB of spectral index $n_\mathrm{t}$. 
The strong degeneracy between $n_\mathrm{t}$ and $r$ is also recovered;
our NG15 analyses result in the following approximate linear relationship: 
\begin{equation}\label{eq:lineprimordial}
    n_\mathrm{t}\simeq -0.13\,\log_{10}r+0.7,
\end{equation}
slightly different from that given by \citet{2023ApJ...951L..11A}.
In addition to the no-KA mode,
our NG15 analysis with a wide prior on $\log_{10}r$
yields the secondary reheating mode
shown in the upper figure of Fig.~\ref{fig:triangle},
similar to what \citet{2023ApJ...951L..11A} found.

The reheating mode is nevertheless absent in the posteriors
resulting from our Bayesian fits to the high-S/N mock data.
Unlike NG15, the NGmock free spectrum has high-S/N data points
at the highest frequencies in the PTA band
that cannot be fit by the GW spectrum in the reheating mode,
since the latter is subject to the UV cutoff at those frequencies, 
as explained in Section~\ref{sec:data}.
Thus, while the reheating mode appears in the NG15 result (despite the UV cutoff)
due to the large uncertainties in the NG15 data at high frequencies,
it is however excluded by NGmock.
Generally, the reheating mode is likely to be disfavored in the future,
as long as a correct UV prescription is implemented.

On the other hand, the analyses of the high-precision NGmock free spectrum
reveal the expected KA vs. no-KA bimodality.
In fact, our \verb|PolyChord| nested sampling finds two posterior clusters
corresponding to the KA and the no-KA modes.
As a result, the 2D contours of the posterior distribution
reconstructed from the NGmock spectrum
in Fig.~\ref{fig:triangle} (blue contours) show two visible peaks in all panels.
The 1D marginalized posteriors of $n_\mathrm{t}$ and $\log_{10}r$
also exhibit the bimodality.
It is important to note that the KA mode is invisible in the NG15 fit
\emph{only} because of the prior volume effect:
the volume of its best-fit region in the parameter space
is significantly smaller than those of the no-KA and the reheating modes.
The reason for this is that the poor sensitivity of the current NG15 data
can accommodate larger fractions of the prior volume in those modes
due to the wider tails of their distributions.
Only when most of those volumes are excluded by the high-S/N mock data
can the KA mode emerge.
In summary, the NGmock spectrum disfavors the reheating mode
but yields a bimodal posterior of the KA and the no-KA modes.

\begin{deluxetable*}{l|C|C|C|C|C|C|C|C}[ht!]
\tablecaption{Parameter best fits and 68\% credible intervals
for the inflationary+SMBHB SGWB model,
based on the nested sampling analyses of the mock free spectra.
\label{tab:mock}}
\tablewidth{0pt}
\tablehead{
\colhead{} & \multicolumn{4}{c}{NGmock (prior on $\log_{10}r$: $[-20,\,0]$)} \vrule & \multicolumn{4}{c}{EPTAmock} \\
\cline{2-9}
\colhead{} & \multicolumn{2}{c}{best-fit values}\vrule & \multicolumn{2}{c}{68\% intervals}\vrule & \multicolumn{2}{c}{best-fit values}\vrule & \multicolumn{2}{c}{68\% intervals} \\
\colhead{} & \colhead{KA} & \colhead{no-KA}\vrule & \colhead{KA} & \colhead{no-KA}\vrule & \colhead{KA} & \colhead{no-KA}\vrule & \multicolumn{2}{c}{Combined}
}
\startdata
$\log_{10}r$ & -4.4 & -11.5 & -5.5^{+3.0}_{-1.2} & -10.2^{+1.6}_{-1.9} & -9.4 & -16.5 & \multicolumn{2}{c}{$-14.4\pm 2.8$} \\
$n_\mathrm{t}$ & 1.01 & 2.22 & 1.29^{+0.13}_{-0.42} & 2.04^{+0.26}_{-0.19} & 1.88 & 2.85 & \multicolumn{2}{c}{$2.58\pm 0.36$} \\
$\log_{10}\kappa_{10}$ & 2.04 & -6.43 & -0.5^{+2.6}_{-1.6} & -5.05^{+0.93}_{-1.5} & -1.79 & -5.85 & \multicolumn{2}{c}{$< -4.05$} \\
$\log_{10}(T_\mathrm{re}/\mathrm{GeV})~~$ & -0.76 & 0.34 & -0.136^{+0.003}_{-0.94} & 3.3^{+2.5}_{-2.3} & -0.36 & 0.23 & \multicolumn{2}{c}{$3.1^{+1.9}_{-2.9}$} \\
$\Delta N_\mathrm{re}$ & 39 & 23 & 24 \pm 20 & 21\pm 10 & 13 & 40 & \multicolumn{2}{c}{---} \\
\hline
$\log_{10}A_\mathrm{BHB}$ & -15.68 & -15.61 & -15.73^{+0.48}_{-0.41} & -15.75^{+0.49}_{-0.40} & -15.60 & -15.58 & \multicolumn{2}{c}{$-15.73^{+0.52}_{-0.41}$} \\
$\gamma_\mathrm{BHB}$ & 4.73 & 4.68 & 4.65\pm 0.33 & 4.65\pm 0.33 & 4.67 & 4.79 & \multicolumn{2}{c}{$4.66\pm 0.35$}
\enddata
\tablecomments{For the EPTAmock analysis result,
the KA and no-KA modes are not separately discernible in the posterior distribution,
so the 68\% interval shown here for that case refers to the combined sample.
}
\end{deluxetable*}

The exclusion of the reheating mode leads to approximately zero likelihood
for regions in the parameter space with $\log_{10}r<-20$.
Therefore, we opt to report the results of the NGmock analyses
obtained using a narrow prior on $\log_{10}r$ for the rest of the paper,
shown in the lower figure of Fig.~\ref{fig:triangle}.
The result obtained by nested sampling
has $n_\mathrm{t} = 1.29^{+0.13}_{-0.42}$ for the KA mode
and $n_\mathrm{t} = 2.04^{+0.26}_{-0.19}$ for the no-KA mode (68\% credible intervals).
This confirms the interpretation that in the KA mode, the PTA data are fit
by a kination-amplified inflationary SGWB of spectral index $n_\mathrm{t}+1$. 
The constraints on all model parameters as well as their best-fit values
are presented in Table~\ref{tab:mock}.

Given the bimodal posterior, we can compare the statistical weights of the two modes
by means of the Bayes factor, as typically done for model selection.
The Bayes factor for the no-KA versus KA mode is
defined as the ratio of their evidences,
$\mathcal{B}\equiv\mathcal{Z}_\mathrm{no-KA}/\mathcal{Z}_\mathrm{KA}$,
such that a large value ($\mathcal{B}>10)$ 
would indicate strong evidence in favor of the no-KA mode.
By contrast, our \verb|PolyChord| analysis yields only $\mathcal{B}=2.9\pm1.9$, 
interpreted as between `negligibly small' and `substantial'
according to the Jeffreys scale \citep{1961Jeffreys}.
Hence, the high-precision NGmock data are 
statistically compatible with (not against) the KA interpretation. 
As a matter of fact, even the result of the corresponding MCMC analysis
exhibit some degree of bimodality, 
indicated by the red contours in Fig.~\ref{fig:triangle}.
However, the MCMC analysis shows quantitative differences from
the one obtained by nested sampling.
Since nested samplers converge more efficiently for multimodal posteriors
and calculate evidences on the fly \citep{2015MNRAS.453.4384H, 2017JCAP...07..033L},
we opt to report the Bayes factor above from the \verb|PolyChord| analysis.

In addition to the Bayes factor,
it is also useful to compare the maximum posterior ratios of the two modes,
$\mathcal{R}\equiv\mathrm{max}\,(\mathcal{P}_\mathrm{no-KA})/
\mathrm{max}\,(\mathcal{P}_\mathrm{KA})$,
associated with the best-fit models listed in Table~\ref{tab:mock} 
(illustrated in Fig.~\ref{fig:spectra}).
We find that $\mathcal{R}=0.8$ for NGmock, 
so the global best fit to the NGmock spectrum is actually given by the KA mode.
This is in contrast to the Bayes factor above, 
which marginally favors the no-KA mode.
Yet no contradiction arises, because what the Bayes factor
or the evidence ratio compares is
the posterior-weighted \emph{volumes} of the two modes.
Furthermore, we will see in Section~\ref{sec:discussion} that 
the Bayes factor defined here depends on the spectral index of the data.
This will explain why the posterior resulting from the EPTAmock spectrum
does not appear to be bimodal (see Fig.~\ref{fig:triangle_EPTA}).
Even if we cannot compute the Bayes factor in this case, 
the global best fit to the EPTAmock spectrum is again provided by the KA mode, 
supporting the existence of bimodality.
The maximum posterior ratio for EPTAmock is $\mathcal{R}=1.0$.

\section{Discussion}\label{sec:discussion}

The above analysis results for the current and mock PTA data imply that
the statistical significance of the secondary KA mode (hence the bimodality)
is strongly dependent on the sensitivity of the experiment.
Existing PTA data cannot decisively exclude
low-likelihood regions (tails of the likelihood) in the parameter space.
As a result, the prior choice can severely influence the marginalized posteriors,
or evidences, for modes that span large prior volumes.
For this reason (the prior volume effect), \citet{2024A&A...685A..94E} 
did not attempt any Bayesian model selection on the origin of the SGWB signal.
The same caveat is known to occur in other fields as well,
such as the interpretation of the Dark Energy Spectroscopic Instrument (DESI)
full-shape clustering data \citep{2024arXiv241112022D}, 
referred to as the ``parameter projection effect'' there.
The DESI team also found that in some cases
the mean of the marginalized posterior can be significantly offset
from the global best-fit value.

Nevertheless, Bayesian model/mode selection should be less subject to the volume effect
as future PTA data become more informative,
i.e., low-likelihood regions will no longer boost
the evidence of modes with larger prior volumes significantly.
Our analysis of the high-precision NGmock spectrum indeed 
demonstrates the quenching of the prior volume effect, 
allowing the bimodality to emerge.
The resulting Bayes factor indicates
comparable statistical significance between the two modes.

\begin{figure}[ht!]
\epsscale{1.2}
\plotone{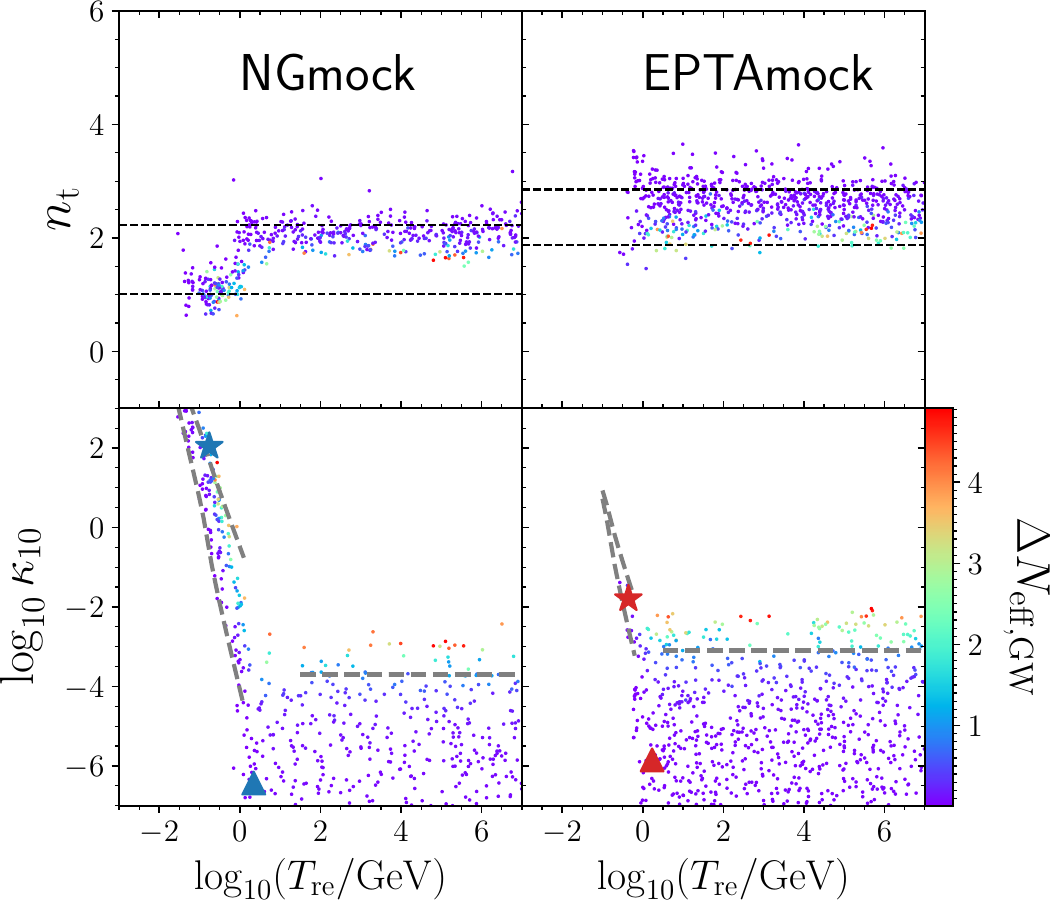}
\caption{3D scatter plot with the NGmock sample (left panels) 
and the EPTAmock sample (right panels).
The color bar indicates the value of $\Delta N_\mathrm{eff,GW}$ for each point.
The stars and the triangles in the lower panels 
denote the KA and the no-KA best-fits to the mock spectra, respectively.
The dashed lines in the upper panels also denote these best fits
(same as in the lower figure of Fig.~\ref{fig:triangle}),
specified in Table~\ref{tab:mock}.
The dashed lines in the lower panels indicate the various physical constraints 
on the KA and the no-KA modes, described by Eqs.~(\ref{eq:wedge_lowerbound}), 
(\ref{eq:wedge_upperbound}) and (\ref{eq:noKA_upperbound}).
}
\label{fig:3D}
\end{figure}

In fact, we are able to identify the physical constraints
that shape the high-likelihood regions in the parameter space for both modes,
which then contribute to the Bayes factor.
In the $\log_{10}\kappa_{10}-\log_{10} (T_\mathrm{re}/\mathrm{GeV})$ panels
of Fig.~\ref{fig:3D} (lower panels),
the KA mode resides in a wedge toward the upper left corner.
The boundaries of this wedge have the following physical interpretations:
\begin{itemize}[leftmargin=10pt,itemindent=0pt]

\item The lower bound of the KA subsample is determined by the requirement that
the peak frequency of the kination-amplified inflationary SGWB
must lie at a value greater than the maximum frequency covered by the PTA band
in order for the KA mode to be a good fit.
Since the peak frequency in the KA mode
corresponds to the frequency of a fluctuation which reentered the horizon
at $T_\mathrm{re}$ \citep{2017PhRvD..96f3505L, 2021JCAP...10..024L},
the above constraint can be expressed as
$f_\mathrm{re}>\mathrm{max}\,(f_\mathrm{PTA})$. It yields
\begin{equation}\label{eq:wedge_lowerbound}
    \log_{10}\kappa_{10} + 4\log_{10}\frac{T_\mathrm{re}}{\mathrm{GeV}}
    +\frac{4}{3}\log_{10}\frac{g_{*,s}(T_\mathrm{re})}{10.75}>-3,
\end{equation}
where $g_{*,s}(T_\mathrm{re})$ is the effective number of 
relativistic degrees of freedom for entropy at $T_\mathrm{re}$.
In this work, the thermal history that determines $g_{*,s}(T)$
is adopted from the tabulated function in \citet{2020JCAP...08..011S} 
for $T\geq10~\mathrm{MeV}$ and computed using the \verb|FortEPiaNO| package
\citep{2019JCAP...07..014G} for $T<10~\mathrm{MeV}$,
which incorporates accurate prescriptions 
for the out-of-equilibrium neutrino decoupling.

\item The upper bound of the KA subsample arises from the constraints
on the effective number of \emph{extra} relativistic species, 
$\Delta N_\mathrm{eff}\equiv N_\mathrm{eff}-N_\mathrm{eff}^\mathrm{SM}$,
where $N_\mathrm{eff}^\mathrm{SM}=3.044$ accounts for
the standard-model neutrinos \citep{2021JCAP...04..073B}.
In this case, the kination-amplified inflationary SGWB 
can contribute a significant $\Delta N_\mathrm{eff}$
\citep{2016PhRvX...6a1035L, 2017PhRvD..96f3505L, 2021JCAP...10..024L}.
While we do not include any $N_\mathrm{eff}$ data in the Bayesian analyses here,
our model is designed to reject points with $\Delta N_\mathrm{eff}>\mathcal{O}(1)$,
in accordance with its current observational constraints from BBN and the CMB
\citep{2022PTEP.2022h3C01W,2020A&A...641A...6P}.
The constraint of $\Delta N_\mathrm{eff,GW}\lesssim\mathcal{O}(1)$ yields
\begin{IEEEeqnarray}{rl}\label{eq:wedge_upperbound}
    & \log_{10}\frac{6}{n_\mathrm{t}+1}
    +\left(10+\frac{2}{3}\log_{10}\frac{g_{*,s}(T_\mathrm{re})}{10.75}\right)n_\mathrm{t}
    +\log_{10}r \nonumber\\
    & + \left(\frac{n_\mathrm{t}}{2}+1\right)\log_{10}\kappa_{10}
    +2(n_\mathrm{t}+1)\,\log_{10}\frac{T_\mathrm{re}}{\mathrm{GeV}} \lesssim 6.
\end{IEEEeqnarray}

\end{itemize}
Eqs.~(\ref{eq:wedge_lowerbound}) and (\ref{eq:wedge_upperbound}) 
set the lower and upper limits of the KA wedge, respectively,
as illustrated in Fig.~\ref{fig:3D}.
Now, let us turn to the no-KA subsample.
It is also subject to the constraint of $\Delta N_\mathrm{eff,GW}\lesssim\mathcal{O}(1)$.
In the no-KA mode, this bound is reached when the inflationary SGWB
is actually kination-amplified at frequencies higher than the PTA band.
The constraint is therefore on the $\kappa_{10}$ parameter 
which determines the transition from kination to the RD era, described by
\begin{IEEEeqnarray}{rl}\label{eq:noKA_upperbound}
    & \log_{10}\frac{4}{n_\mathrm{t}+1}+\frac{8.5-\log_{10}r}{n_\mathrm{t}}
    +\frac{1}{2}\log_{10}\kappa_{10} \nonumber\\
    & +\frac{1}{3}\log_{10}\frac{g_{*,s}(T_\mathrm{kr})}{10.75} \lesssim 8,
    \qquad 
    T_\mathrm{kr}\approx \frac{10\,\mathrm{MeV}}{\sqrt{\kappa_{10}}},\quad
\end{IEEEeqnarray}
where $T_\mathrm{kr}$ is the temperature at kination-radiation equality.
Eq.~(\ref{eq:noKA_upperbound}) is illustrated by the horizontal dashed lines
in the lower panels of Fig.~\ref{fig:3D}.

The above physical constraints place hard bounds on the volumes in the parameter space
that contribute to the evidences of the modes.
They hence impact the relative statistical significance
between the no-KA mode and the KA mode, or the Bayes factor.
The lower panels of Fig.~\ref{fig:3D} show that
the KA wedge from EPTAmock is much smaller than that from NGmock,
explaining why the EPTAmock sample fails to display the KA mode.
Furthermore, we can infer from Eqs.~(\ref{eq:wedge_upperbound}) 
and (\ref{eq:noKA_upperbound}) that the volumes of the constrained regions
depend on $n_\mathrm{t}$.\footnote{Since all good fits approximately satisfy 
the tight relationship between $\log_{10}r$ and $n_\mathrm{t}$
in Eq.~(\ref{eq:lineprimordial}),
$n_\mathrm{t}$ is effectively the only variable 
in the constraints in Equations~(\ref{eq:wedge_upperbound}) and (\ref{eq:noKA_upperbound}).
} 
In fact, the smaller KA wedge from the EPTAmock spectra
results exactly from its higher best-fit value of $n_\mathrm{t}$.
Thus, the index of the free spectrum measured by PTAs
affects the Bayes factor for the no-KA~vs.~KA mode comparison,
as mentioned in Section~\ref{sec:results}.

For given PTA data sets, the KA interpretation has the advantage of
demanding a less blue-tilted primordial tensor spectrum,
compared with the ordinary, no-KA mode 
(i.e., $n_\mathrm{t}+1$~vs.~$n_\mathrm{t}$ for the spectral index
in $\Omega_\mathrm{GW}$, the SGWB spectrum today).
This is helpful since (i) plausible early-Universe scenarios
that predict large blue tensor tilts ($n_\mathrm{t}>1$) 
and satisfy the $N_\mathrm{eff}$ constraint at the same time
are difficult to realize \citep{2023MNRAS.520.1757G},
and (ii) the KA interpretation of the PTA signal
is less subject to the UV cutoff than the no-KA interpretation
(i.e., because it occurs at higher frequencies).
Therefore, the kination-amplified inflationary SGWB here does not only
present a case study of the multimodality in new-physics searches,
but serves as an important model for the origin of the PTA signal on its own.

\section{Conclusions and outlook}\label{sec:conclusion}

In this paper, we have investigated the kination-amplified inflationary SGWB
interpretation of the recently reported GW signal from various PTA experiments.
Kination refers to a post-inflationary, pre-BBN phase
in the expansion history dominated by the kinetic energy of some scalar field,
thus characterized by the EoS parameter of $w=1$.
Within the inflationary SGWB model,
we have identified two modes which can fit the PTA data sets with equal likelihood:
the kination-amplification (KA) mode and the ordinary, no-KA mode.
Motivated by the possibility of multimodal posteriors,
we have performed Bayesian searches using nested sampling
in both the current PTA data
and futuristic mock data constructed with higher S/Ns.
The analysis of the mock free spectrum designed to be consistent with
the best fit to the NANOGrav 15 yr data successfully finds
the expected bimodal posterior for the first time and,
in the meantime, excludes the reheating mode
that appears in fit to the current NG15 data,
making a case for our correct and comprehensive treatment
of potential multimodal posteriors arising from future PTA data sets.
The resultant Bayes factor shows comparable significance
between the KA and the no-KA modes.
While correct modeling of the SGWB signal
(e.g., taking account of the UV cutoff) is critical,
it is also necessary to seek the optimal sampler for multimodal posteriors.
For this purpose, nested-sampling methods should be considered for future PTA analyses,
since they can explore multimodal posteriors more efficiently
than the traditional MCMC methods.

Finally, we provide a few comments on the implications of future data sets
for multimodal posteriors.
Although the mock free spectra used in this work are not realistic,
they mimic the situation in which the sensitive frequency band of the PTAs
widens at the higher end in the future.
In the case of the inflationary SGWB with possible kination amplification,
more informative PTA data at high frequencies with the same spectral index
may further disfavor the no-KA mode.
This will happen if the maximum PTA frequency becomes higher than
the nonlinearity cutoff frequency (detailed in Appendix~\ref{app:model}) 
for the no-KA mode, so that the no-KA mode cannot be a good fit.
By contrast, the KA mode, with a less blue-tilted primordial spectrum,
is less susceptible to the nonlinearity cutoff, as shown in Fig.~\ref{fig:spectra}.
Therefore, the result of mode selection may shift
as future PTA data sets become available.

Moreover, the best-fit spectra
of the inflationary SGWB today shown in Fig.~\ref{fig:spectra} 
encourage the synergy between PTA and other GW data sets. 
On the high-frequency side, the looser UV cutoff for the KA mode 
allows for the associated $\Omega_\mathrm{GW}(f)$ to extend into the mHz frequency band
relevant for the LISA-type, space-borne GW detectors
\citep{2023LRR....26....5A,2020NatAs...4..108R,2021NatAs...5..881G},
whereas the single-power-law, no-KA mode may not allow a PTA+LISA joint fit.
On the low-frequency side, the less blue-tilted primordial spectrum for the KA mode
is also reflected in the shallower lever arm between the CMB scale
and the PTA band in $\Omega_\mathrm{GW}(f)$ \citep{2015PhRvD..91j3505M}.
Therefore, unlike the no-KA mode, the best-fit value of $r$ for the KA mode
could potentially be within reach at the design sensitivity
of the next-generation CMB polarization experiments
searching for primordial gravitational waves \citep{2016arXiv161002743A},
allowing for a CMB+PTA joint fit.
Such joint Bayesian analyses will be accelerated by our recently released
neural network emulator, \verb|SageNet|
\citep[Stiff-Amplified Gravitational-wave Emulator Network,][]{2025arXiv250404054Z}\footnote{Code available at: \url{https://github.com/YifangLuo/SageNet}},
which can efficiently generate the present-day GW spectra
here calculated by \verb|stiffGWpy|.
In summary, a joint analysis of future CMB polarization,
PTA and laser interferometer data will help shed light on models of new physics
that may explain observations of the SGWB,
and will allow for a quantitative comparison of the various physical mechanisms
consistent with the data that give rise to multimodal posteriors
from PTA-only searches.


\begin{acknowledgments}
We thank Joseph Romano, Kejia Lee, Siyuan Chen, Zucheng Chen
for helpful discussions.
BL is supported by the National Natural Science Foundation of China
(Grant Nos. 12203012, 12494575) 
and Guangxi Natural Science Foundation (Grant No. 2023GXNSFBA026114).
This work is also supported by the Guangxi Talent Program
(``Highland of Innovation Talents'').
JM is supported by the US~Department of Energy under Grant~\mbox{DE-SC0010129} and by NASA through Grant~\mbox{80NSSC24K0665}.
PRS acknowledges support from NASA under Grant No. 80NSSC22K175.
\end{acknowledgments}

\software{cobaya \citep{2019ascl.soft10019T, 2021JCAP...05..057T},
          PolyChord \citep{2015MNRAS.450L..61H, 2015MNRAS.453.4384H},
          FortEPiaNO \citep{2019JCAP...07..014G},
          stiffGWpy (\url{https://github.com/bohuarolandli/stiffGWpy}),
          }


\appendix

\section{Details of the model} \label{app:model}

\begin{figure*}[ht!]
\resizebox{\textwidth}{!}{\includegraphics[width=1.5\textwidth]{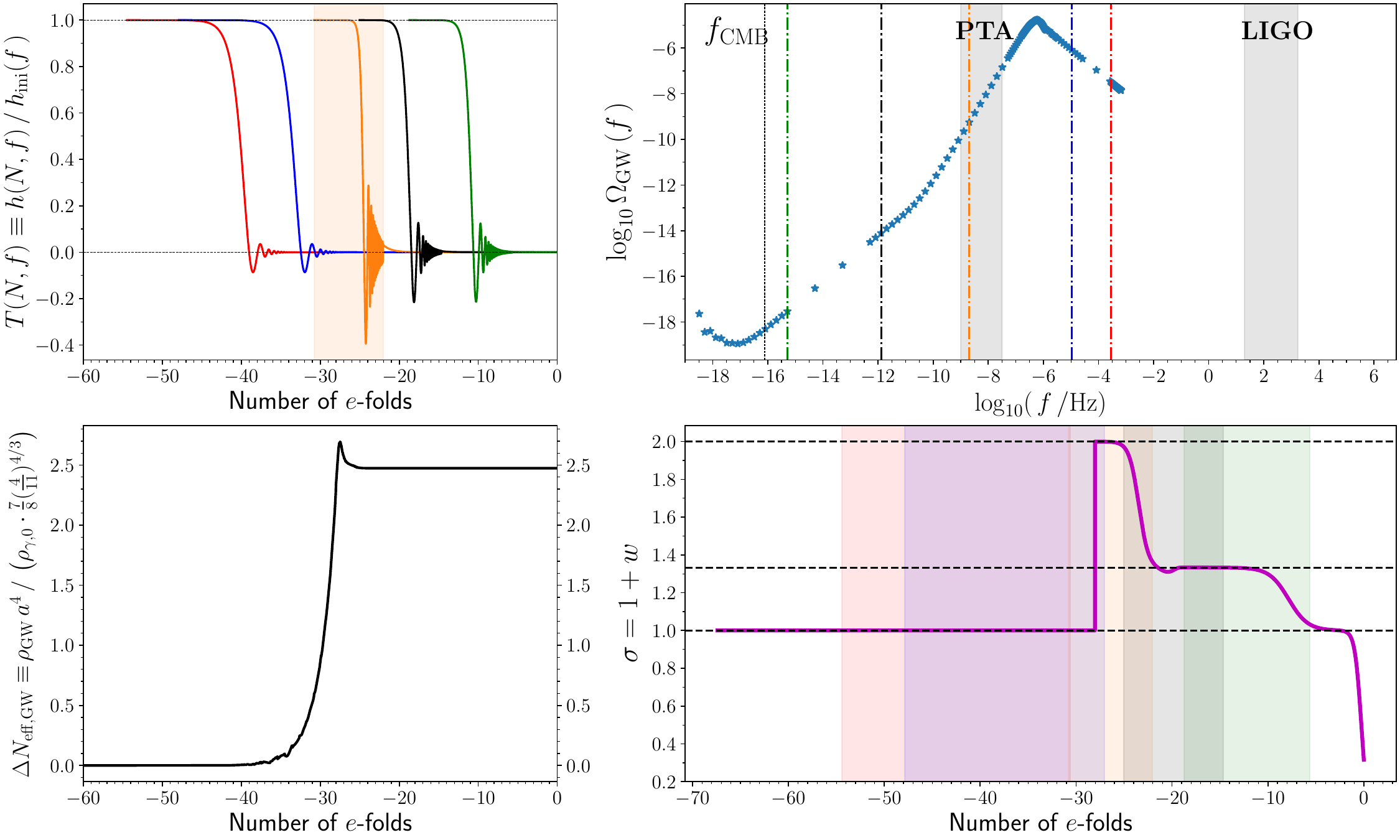}}
\caption{\emph{Upper left}: tensor transfer functions for five example frequencies.
The orange vertical shade indicates the horizon-crossing epoch
of the corresponding frequency.
\emph{Upper right}: Kination-amplified inflationary SGWB spectrum today,
same as that indicated
by the solid curves in the upper panels of Fig.~\ref{fig:spectra}.
The vertical dash-dotted lines denote the example frequencies
illustrated in the upper left panel.
The gray vertical shades indicate the sensitive frequency band of PTAs
and that of Advanced LIGO/Virgo \citep{2021PhRvD.104b2004A}, respectively.
\emph{Lower left}: evolution of the effective number of extra relativistic species
due to the inflationary SGWB, $\Delta N_\mathrm{eff,GW}$.
\emph{Lower right}: expansion history described by the evolution of the EoS parameter.
The vertical shades indicate the horizon-crossing epochs
of the frequencies in the upper left panel.
}
\label{fig:solution}
\end{figure*}

Consistent modeling of the blue-tilted inflationary SGWB with possible kination amplification
involves the treatment of two more complexities:
\begin{itemize}
    \item[(i)] The inflationary SGWB requires a UV cutoff,
    typically chosen as $f_\mathrm{inf}$, 
    corresponding to the horizon scale at the end of inflation.
    However, when $n_\mathrm{t}>0$, it is possible that 
    the amplitude of the tensor modes in the blue-tilted spectrum 
    reaches unity at some high frequency \emph{below} the $f_\mathrm{inf}$ cutoff.
    When this apparent nonlinearity happens, the amplitude must saturate, 
    and one needs a more advanced prescription than linear perturbation theory
    to treat the UV behavior of the primordial tensor fluctuations
    \citep[e.g.,][]{2024JHEP...02..008Y,2024PhRvD.110j3529P}. 
    This pathology is unfortunately overlooked in existing studies
    which attempt to fit inflationary SGWB models to PTA data, 
    e.g., \citet{2023ApJ...951L..11A}.
    
    To avoid dealing with the unknown physics beyond the apparent nonlinearity, 
    we keep the single-power-law assumption
    while imposing another UV cutoff for the primordial tensor spectrum,
    $f_\mathrm{cut}$, at which the tensor power equals unity, 
    $A_\mathrm{t}\,(f_\mathrm{cut}/f_\mathrm{CMB})^{n_\mathrm{t}}\equiv 1$.
    This cutoff is illustrated in the insets of Fig.~\ref{fig:spectra}.
    
    \item[(ii)] As we showed in \citet{2021JCAP...10..024L}, SGWBs with spectra that are blue-tilted (either because the primordial tensor spectrum is blue-tilted 
    or because kination amplification tilts a flat primordial spectrum blue for wavelengths that entered the horizon after reheating) 
    can make a considerable contribution
    to the critical density during radiation domination, 
    leading to a nonnegligible backreaction on the expansion history.
    We account for this backreaction by an iterative numerical scheme
    developed in \citet{2021JCAP...10..024L}; see also \citet{2022MNRAS.509.1366K}.

    As mentioned in Section~\ref{sec:kination}, we solve
    the exact tensor wave equation for a range of sampled frequencies, 
    using a dynamical system approach.
    The tensor wave equation is only integrated for the horizon-crossing epoch
    of each frequency, $2\pi f/aH \in [10^{-3}, e^5]$.
    After it is well inside the horizon, we switch to the time-averaged solution, 
    as demonstrated in the upper left panel of Fig.~\ref{fig:solution}.
    The overall numerical solver for the inflationary SGWB
    with possible kination amplification is released as the \verb|stiffGWpy| package. Fig.~\ref{fig:solution} shows the same example numerical solution
    as in Fig.~\ref{fig:spectra}.
\end{itemize}

Apart from the inflationary SGWB, 
the SGWB from SMBHBs must be taken into account.
Following \citet{2023ApJ...951L..11A}, 
we adopt a power-law model for the SMBHB component, 
parameterized by $(\log_{10} A_\mathrm{BHB},~\gamma_\mathrm{BHB})$.
This pair of parameters approximately satisfies a bivariate Gaussian distribution 
with the following mean and covariance matrix:
\begin{equation}
    \mu = (-15.6,\,4.7),\qquad \sigma = 
    \begin{bmatrix}
        0.28 & -0.0026 \\
        -0.0026 & 0.12
    \end{bmatrix}.
\end{equation}
Moreover, the \verb|PolyChord| nested sampler requires bounded priors,
so we impose the following bounded intervals on these parameters:
$\log_{10} A_\mathrm{BHB}\in[-18,\,-11]$ and $\gamma_\mathrm{BHB}\in[0,\,7]$,
in addition to the bivariate Gaussian prior.


\bibliography{SGWB}{}

\begin{thebibliography}{}
\expandafter\ifx\csname natexlab\endcsname\relax\def\natexlab#1{#1}\fi
\providecommand{\url}[1]{\href{#1}{#1}}
\providecommand{\dodoi}[1]{doi:~\href{http://doi.org/#1}{\nolinkurl{#1}}}
\providecommand{\doeprint}[1]{\href{http://ascl.net/#1}{\nolinkurl{http://ascl.net/#1}}}
\providecommand{\doarXiv}[1]{\href{https://arxiv.org/abs/#1}{\nolinkurl{https://arxiv.org/abs/#1}}}

\bibitem[{K.~N. {Abazajian} {et~al.}(2016){Abazajian}, {Adshead}, {Ahmed},
  {Allen}, {Alonso}, {Arnold}, {Baccigalupi}, {Bartlett}, {Battaglia},
  {Benson}, {Bischoff}, {Borrill}, {Buza}, {Calabrese}, {Caldwell},
  {Carlstrom}, {Chang}, {Crawford}, {Cyr-Racine}, {De Bernardis}, {de Haan},
  {di Serego Alighieri}, {Dunkley}, {Dvorkin}, {Errard}, {Fabbian}, {Feeney},
  {Ferraro}, {Filippini}, {Flauger}, {Fuller}, {Gluscevic}, {Green}, {Grin},
  {Grohs}, {Henning}, {Hill}, {Hlozek}, {Holder}, {Holzapfel}, {Hu},
  {Huffenberger}, {Keskitalo}, {Knox}, {Kosowsky}, {Kovac}, {Kovetz}, {Kuo},
  {Kusaka}, {Le Jeune}, {Lee}, {Lilley}, {Loverde}, {Madhavacheril}, {Mantz},
  {Marsh}, {McMahon}, {Meerburg}, {Meyers}, {Miller}, {Munoz}, {Nguyen},
  {Niemack}, {Peloso}, {Peloton}, {Pogosian}, {Pryke}, {Raveri}, {Reichardt},
  {Rocha}, {Rotti}, {Schaan}, {Schmittfull}, {Scott}, {Sehgal}, {Shandera},
  {Sherwin}, {Smith}, {Sorbo}, {Starkman}, {Story}, {van Engelen}, {Vieira},
  {Watson}, {Whitehorn}, \& {Kimmy Wu}}]{2016arXiv161002743A}
{Abazajian}, K.~N., {Adshead}, P., {Ahmed}, Z., {et~al.} 2016,
  \bibinfo{title}{{CMB-S4 Science Book, First Edition},} arXiv e-prints,
  arXiv:1610.02743.
\newblock \doarXiv{1610.02743}

\bibitem[{L.~F. {Abbott} \& M.~B. {Wise}(1984){Abbott} \&
  {Wise}}]{1984NuPhB.244..541A}
{Abbott}, L.~F., \& {Wise}, M.~B. 1984, \bibinfo{title}{{Constraints on
  generalized inflationary cosmologies},} Nuclear Physics B, 244, 541,
  \dodoi{10.1016/0550-3213(84)90329-8}

\bibitem[{R. {Abbott} {et~al.}(2021){Abbott}, {Abbott}, {Abraham}, {Acernese},
  {Ackley}, {Adams}, {Adams}, {Adhikari}, {Adya}, {Affeldt}, {Agarwal},
  {Agathos}, {Agatsuma}, {Aggarwal}, {Aguiar}, {Aiello}, {Ain}, {Akutsu},
  {Aleman}, {Allen}, {Allocca}, {Altin}, {Amato}, {Anand}, {Ananyeva},
  {Anderson}, {Anderson}, {Ando}, {Angelova}, {Ansoldi}, {Antelis}, {Antier},
  {Appert}, {Arai}, {Arai}, {Arai}, {Araki}, {Araya}, {Araya}, {Areeda},
  {Ar{\`e}ne}, {Aritomi}, {Arnaud}, {Aronson}, {Asada}, {Asali}, {Ashton},
  {Aso}, {Aston}, {Astone}, {Aubin}, {Aufmuth}, {Aultoneal}, {Austin}, {Babak},
  {Badaracco}, {Bader}, {Bae}, {Bae}, {Baer}, {Bagnasco}, {Bai}, {Baiotti},
  {Baird}, {Bajpai}, {Ball}, {Ballardin}, {Ballmer}, {Bals}, {Balsamo},
  {Baltus}, {Banagiri}, {Bankar}, {Bankar}, {Barayoga}, {Barbieri}, {Barish},
  {Barker}, {Barneo}, {Barnum}, {Barone}, {Barr}, {Barsotti}, {Barsuglia},
  {Barta}, {Bartlett}, {Barton}, {Bartos}, {Bassiri}, {Basti}, {Bawaj},
  {Bayley}, {Baylor}, {Bazzan}, {B{\'e}csy}, {Bedakihale}, {Bejger},
  {Belahcene}, {Benedetto}, {Beniwal}, {Benjamin}, {Bennett}, {Bentley},
  {Benyaala}, {Bergamin}, {Berger}, {Bernuzzi}, {Bersanetti}, {Bertolini},
  {Betzwieser}, {Bhandare}, {Bhandari}, {Bhattacharjee}, {Bhaumik}, {Bidler},
  {Bilenko}, {Billingsley}, {Birney}, {Birnholtz}, {Biscans}, {Bischi},
  {Biscoveanu}, {Bisht}, {Biswas}, {Bitossi}, {Bizouard}, {Blackburn},
  {Blackman}, {Blair}, {Blair}, {Blair}, {Bobba}, {Bode}, {Boer}, {Bogaert},
  {Boldrini}, {Bondu}, {Bonilla}, {Bonnand}, {Booker}, {Boom}, {Bork},
  {Boschi}, {Bose}, {Bose}, {Bossilkov}, {Boudart}, {Bouffanais}, {Bozzi},
  {Bradaschia}, {Brady}, {Bramley}, {Branch}, {Branchesi}, {Brau}, {Breschi},
  {Briant}, {Briggs}, {Brillet}, {Brinkmann}, {Brockill}, {Brooks}, {Brooks},
  {Brown}, {Brunett}, {Bruno}, {Bruntz}, {Bryant}, {Buikema}, {Bulik},
  {Bulten}, {Buonanno}, {Buscicchio}, {Buskulic}, {Byer}, {Cadonati}, {Caesar},
  {Cagnoli}, {Cahillane}, {Cain}, {Bustillo}, {Callaghan}, {Callister},
  {Calloni}, {Camp}, {Canepa}, {Cannavacciuolo}, {Cannon}, {Cao}, {Cao}, {Cao},
  {Capocasa}, {Capote}, {Carapella}, {Carbognani}, {Carlin}, {Carney},
  {Carpinelli}, {Carullo}, {Carver}, {Diaz}, {Casentini}, {Castaldi},
  {Caudill}, {Cavagli{\`a}}, {Cavalier}, {Cavalieri}, {Cella},
  {Cerd{\'a}-Dur{\'a}n}, {Cesarini}, {Chaibi}, {Chakravarti}, {Champion},
  {Chan}, {Chan}, {Chan}, {Chan}, {Chandra}, {Chanial}, {Chao}, {Charlton},
  {Chase}, {Chassande-Mottin}, {Chatterjee}, {Chaturvedi}, {Chen}, {Chen},
  {Chen}, {Chen}, {Chen}, {Chen}, {Chen}, {Chen}, {Chen}, {Cheng}, {Cheong},
  {Cheung}, {Chia}, {Chiadini}, {Chiang}, {Chierici}, {Chincarini}, {Chiofalo},
  {Chiummo}, {Cho}, {Cho}, {Choate}, {Choudhary}, {Choudhary}, {Christensen},
  {Chu}, {Chu}, {Chu}, {Chua}, {Chung}, {Ciani}, {Ciecielag}, {Cie{\'s}lar},
  {Cifaldi}, {Ciobanu}, {Ciolfi}, {Cipriano}, {Cirone}, {Clara}, {Clark},
  {Clark}, {Clarke}, {Clearwater}, {Clesse}, {Cleva}, {Coccia}, {Cohadon},
  {Cohen}, {Cohen}, {Colleoni}, {Collette}, {Colpi}, {Compton}, {Constancio},
  {Conti}, {Cooper}, {Corban}, {Corbitt}, {Cordero-Carri{\'o}n}, {Corezzi},
  {Corley}, {Cornish}, {Corre}, {Corsi}, {Cortese}, {Costa}, {Cotesta},
  {Coughlin}, {Coughlin}, {Coulon}, {Countryman}, {Cousins}, {Couvares},
  {Covas}, {Coward}, {Cowart}, {Coyne}, {Coyne}, {Creighton}, {Creighton},
  {Criswell}, {Croquette}, {Crowder}, {Cudell}, {Cullen}, {Cumming},
  {Cummings}, {Cuoco}, {Cury{\l}o}, {Canton}, {D{\'a}lya}, {Dana},
  {Daneshgaranbajastani}, {D'Angelo}, {Danilishin}, {D'Antonio}, {Danzmann},
  {Darsow-Fromm}, {Dasgupta}, {Datrier}, {Dattilo}, {Dave}, {Davier}, {Davies},
  {Davis}, {Daw}, {Dean}, {Deenadayalan}, {Degallaix}, {de Laurentis},
  {Del{\'e}glise}, {Del Favero}, {de Lillo}, {de Lillo}, {Del Pozzo},
  {Demarchi}, {de Matteis}, {D'Emilio}, {Demos}, {Dent}, {Depasse}, {de
  Pietri}, {De Rosa}, {de Rossi}, {Desalvo}, {de Simone}, {Dhurandhar},
  {D{\'\i}az}, {Diaz-Ortiz}, {Didio}, {Dietrich}, {di Fiore}, {di Fronzo}, {di
  Giorgio}, {di Giovanni}, {di Girolamo}, {di Lieto}, {Ding}, {di Pace}, {di
  Palma}, {di Renzo}, {Divakarla}, {Dmitriev}, {Doctor}, {D'Onofrio},
  {Donovan}, {Dooley}, {Doravari}, {Dorrington}, {Drago}, {Driggers}, {Drori},
  {Du}, {Ducoin}, {Dupej}, {Durante}, {D'Urso}, {Duverne}, {Dvorkin}, {Dwyer},
  {Easter}, {Ebersold}, {Eddolls}, {Edelman}, {Edo}, {Edy}, {Effler}, {Eguchi},
  {Eichholz}, {Eikenberry}, {Eisenmann}, {Eisenstein}, {Ejlli}, {Enomoto},
  {Errico}, {Essick}, {Estell{\'e}s}, {Estevez}, {Etienne}, {Etzel}, {Evans},
  {Evans}, {Ewing}, {Fafone}, {Fair}, {Fairhurst}, {Fan}, {Farah}, {Farinon},
  {Farr}, {Farr}, {Farrow}, {Fauchon-Jones}, {Favata}, {Fays}, {Fazio},
  {Feicht}, {Fejer}, {Feng}, {Fenyvesi}, {Ferguson}, {Fernandez-Galiana},
  {Ferrante}, {Ferreira}, {Fidecaro}, {Figura}, {Fiori}, {Fishbach}, {Fisher},
  {Fishner}, {Fittipaldi}, {Fiumara}, {Flaminio}, {Floden}, {Flynn}, {Fong},
  {Font}, {Fornal}, {Forsyth}, {Franke}, {Frasca}, {Frasconi}, {Frederick},
  {Frei}, {Freise}, {Frey}, {Fritschel}, {Frolov}, {Fronz{\'e}}, {Fujii},
  {Fujikawa}, {Fukunaga}, {Fukushima}, {Fulda}, {Fyffe}, {Gabbard}, {Gadre},
  {Gaebel}, {Gair}, {Gais}, {Galaudage}, {Gamba}, {Ganapathy}, {Ganguly},
  {Gao}, {Gaonkar}, {Garaventa}, {Garc{\'\i}a-N{\'u}{\~n}ez},
  {Garc{\'\i}a-Quir{\'o}s}, {Garufi}, {Gateley}, {Gaudio}, {Gayathri}, {Ge},
  {Gemme}, {Gennai}, {George}, {Gergely}, {Gewecke}, {Ghonge}, {Ghosh},
  {Ghosh}, {Ghosh}, {Ghosh}, {Ghosh}, {Giacomazzo}, {Giacoppo}, {Giaime},
  {Giardina}, {Gibson}, {Gier}, {Giesler}, {Giri}, {Gissi}, {Glanzer},
  {Gleckl}, {Godwin}, {Goetz}, {Goetz}, {Gohlke}, {Goncharov}, {Gonz{\'a}lez},
  {Gopakumar}, {Gosselin}, {Gouaty}, {Grace}, {Grado}, {Granata}, {Granata},
  {Grant}, {Gras}, {Grassia}, {Gray}, {Gray}, {Greco}, {Green}, {Green},
  {Gretarsson}, {Gretarsson}, {Griffith}, {Griffiths}, {Griggs}, {Grignani},
  {Grimaldi}, {Grimes}, {Grimm}, {Grote}, {Grunewald}, {Gruning}, {Guerrero},
  {Guidi}, {Guimaraes}, {Guix{\'e}}, {Gulati}, {Guo}, {Guo}, {Gupta}, {Gupta},
  {Gupta}, {Gustafson}, {Gustafson}, {Guzman}, {Ha}, {Haegel}, {Hagiwara},
  {Haino}, {Halim}, {Hall}, {Hamilton}, {Hammond}, {Han}, {Haney}, {Hanks},
  {Hanna}, {Hannam}, {Hannuksela}, {Hansen}, {Hansen}, {Hanson}, {Harder},
  {Hardwick}, {Haris}, {Harms}, {Harry}, {Harry}, {Hartwig}, {Hasegawa},
  {Haskell}, {Hasskew}, {Haster}, {Hattori}, {Haughian}, {Hayakawa}, {Hayama},
  {Hayes}, {Healy}, {Heidmann}, {Heintze}, {Heinze}, {Heinzel}, {Heitmann},
  {Hellman}, {Hello}, {Helmling-Cornell}, {Hemming}, {Hendry}, {Heng},
  {Hennes}, {Hennig}, {Hennig}, {Vivanco}, {Heurs}, {Hild}, {Hill}, {Himemoto},
  {Hines}, {Hiranuma}, {Hirata}, {Hirose}, {Hochheim}, {Hofman}, {Hohmann},
  {Holgado}, {Holland}, {Hollows}, {Holmes}, {Holt}, {Holz}, {Hong}, {Hopkins},
  {Hough}, {Howell}, {Hoy}, {Hoyland}, {Hreibi}, {Hsieh}, {Hsu}, {Huang},
  {Huang}, {Huang}, {Huang}, {Huang}, {Huang}, {H{\"u}bner}, {Huddart},
  {Huerta}, {Hughey}, {Hui}, {Hui}, {Husa}, {Huttner}, {Huxford}, {Huynh-Dinh},
  {Ide}, {Idzkowski}, {Iess}, {Ikenoue}, {Imam}, {Inayoshi}, {Inchauspe},
  {Ingram}, {Inoue}, {Intini}, {Ioka}, {Isi}, {Isleif}, {Ito}, {Itoh}, {Iyer},
  {Izumi}, {Jaberianhamedan}, {Jacqmin}, {Jadhav}, {Jadhav}, {James}, {Jan},
  {Jani}, {Janssens}, {Janthalur}, {Jaranowski}, {Jariwala}, {Jaume},
  {Jenkins}, {Jeon}, {Jeunon}, {Jia}, {Jiang}, {Jin}, {Johns}, {Jones},
  {Jones}, {Jones}, {Jones}, {Jones}, {Jonker}, {Ju}, {Jung}, {Jung}, {Junker},
  {Kaihotsu}, {Kajita}, {Kakizaki}, {Kalaghatgi}, {Kalogera}, {Kamai},
  {Kamiizumi}, {Kanda}, {Kandhasamy}, {Kang}, {Kanner}, {Kao}, {Kapadia},
  {Kapasi}, {Karathanasis}, {Karki}, {Kashyap}, {Kasprzack}, {Kastaun},
  {Katsanevas}, {Katsavounidis}, {Katzman}, {Kaur}, {Kawabe}, {Kawaguchi},
  {Kawai}, {Kawasaki}, {K{\'e}f{\'e}lian}, {Keitel}, {Key}, {Khadka},
  {Khalili}, {Khan}, {Khan}, {Khazanov}, {Khetan}, {Khursheed}, {Kijbunchoo},
  {Kim}, {Kim}, {Kim}, {Kim}, {Kim}, {Kim}, {Kimball}, {Kimura}, {King},
  {Kinley-Hanlon}, {Kirchhoff}, {Kissel}, {Kita}, {Kitazawa}, {Kleybolte},
  {Klimenko}, {Knee}, {Knowles}, {Knyazev}, {Koch}, {Koekoek}, {Kojima},
  {Kokeyama}, {Koley}, {Kolitsidou}, {Kolstein}, {Komori}, {Kondrashov},
  {Kong}, {Kontos}, {Koper}, {Korobko}, {Kotake}, {Kovalam}, {Kozak},
  {Kozakai}, {Kozu}, {Kringel}, {Krishnendu}, {Kr{\'o}lak}, {Kuehn}, {Kuei},
  {Kumar}, {Kumar}, {Kumar}, {Kumar}, {Kume}, {Kuns}, {Kuo}, {Kuo}, {Kuromiya},
  {Kuroyanagi}, {Kusayanagi}, {Kwak}, {Kwang}, {Laghi}, {Lalande}, {Lam},
  {Lamberts}, {Landry}, {Lane}, {Lang}, {Lange}, {Lantz}, {La Rosa},
  {Lartaux-Vollard}, {Lasky}, {Laxen}, {Lazzarini}, {Lazzaro}, {Leaci},
  {Leavey}, {Lecoeuche}, {Lee}, {Lee}, {Lee}, {Lee}, {Lee}, {Lee}, {Lehmann},
  {Lema{\^\i}tre}, {Leon}, {Leonardi}, {Leroy}, {Letendre}, {Levin}, {Leviton},
  {Li}, {Li}, {Li}, {Li}, {Li}, {Li}, {Lin}, {Lin}, {Lin}, {Lin}, {Lin},
  {Linde}, {Linker}, {Linley}, {Littenberg}, {Liu}, {Liu}, {Liu}, {Liu},
  {Llorens-Monteagudo}, {Lo}, {Lockwood}, {Lollie}, {London}, {Longo}, {Lopez},
  {Lorenzini}, {Loriette}, {Lormand}, {Losurdo}, {Lough}, {Lousto}, {Lovelace},
  {L{\"u}ck}, {Lumaca}, {Lundgren}, {Luo}, {Macas}, {Macinnis}, {MacLeod},
  {MacMillan}, {Macquet}, {Hernandez}, {Maga{\~n}a-Sandoval}, {Magazz{\`u}},
  {Magee}, {Maggiore}, {Majorana}, {Maksimovic}, {Maliakal}, {Malik}, {Man},
  {Mandic}, {Mangano}, {Mango}, {Mansell}, {Manske}, {Mantovani}, {Mapelli},
  {Marchesoni}, {Marchio}, {Marion}, {Mark}, {M{\'a}rka}, {M{\'a}rka},
  {Markakis}, {Markosyan}, {Markowitz}, {Maros}, {Marquina}, {Marsat},
  {Martelli}, {Martin}, {Martin}, {Martinez}, {Martinez}, {Martinovic},
  {Martynov}, {Marx}, {Masalehdan}, {Mason}, {Massera}, {Masserot},
  {Massinger}, {Masso-Reid}, {Mastrogiovanni}, {Matas}, {Mateu-Lucena},
  {Matichard}, {Matiushechkina}, {Mavalvala}, {McCann}, {McCarthy},
  {McClelland}, {McClincy}, {McCormick}, {McCuller}, {McGhee}, {McGuire},
  {McIsaac}, {McIver}, {McManus}, {McRae}, {McWilliams}, {Meacher}, {Mehmet},
  {Mehta}, {Melatos}, {Melchor}, {Mendell}, {Menendez-Vazquez}, {Menoni},
  {Mercer}, {Mereni}, {Merfeld}, {Merilh}, {Merritt}, {Merzougui}, {Meshkov},
  {Messenger}, {Messick}, {Meyers}, {Meylahn}, {Mhaske}, {Miani}, {Miao},
  {Michaloliakos}, {Michel}, {Michimura}, {Middleton}, {Milano}, {Miller},
  {Millhouse}, {Mills}, {Milotti}, {Milovich-Goff}, {Minazzoli}, {Minenkov},
  {Mio}, {Mir}, {Mishkin}, {Mishra}, {Mishra}, {Mistry}, {Mitra}, {Mitrofanov},
  {Mitselmakher}, {Mittleman}, {Miyakawa}, {Miyamoto}, {Miyazaki}, {Miyo},
  {Miyoki}, {Mo}, {Mogushi}, {Mohapatra}, {Mohite}, {Molina}, {Molina-Ruiz},
  {Mondin}, {Montani}, {Moore}, {Moraru}, {Morawski}, {More}, {Moreno},
  {Moreno}, {Mori}, {Morisaki}, {Moriwaki}, {Mours}, {Mow-Lowry}, {Mozzon},
  {Muciaccia}, {Mukherjee}, {Mukherjee}, {Mukherjee}, {Mukherjee}, {Mukund},
  {Mullavey}, {Munch}, {Mu{\~n}iz}, {Murray}, {Musenich}, {Nadji}, {Nagano},
  {Nagano}, {Nagar}, {Nakamura}, {Nakano}, {Nakano}, {Nakashima}, {Nakayama},
  {Nardecchia}, {Narikawa}, {Naticchioni}, {Nayak}, {Nayak}, {Negishi}, {Neil},
  {Neilson}, {Nelemans}, {Nelson}, {Nery}, {Neunzert}, {Ng}, {Ng}, {Nguyen},
  {Nguyen}, {Nguyen}, {Quynh}, {Ni}, {Nichols}, {Nishizawa}, {Nissanke},
  {Nocera}, {Noh}, {Norman}, {North}, {Nozaki}, {Nuttall}, {Oberling},
  {O'Brien}, {Obuchi}, {O'Dell}, {Ogaki}, {Oganesyan}, {Oh}, {Oh}, {Oh},
  {Ohashi}, {Ohishi}, {Ohkawa}, {Ohme}, {Ohta}, {Okada}, {Okutani}, {Okutomi},
  {Olivetto}, {Oohara}, {Ooi}, {Oram}, {O'Reilly}, {Ormiston}, {Ormsby},
  {Ortega}, {O'Shaughnessy}, {O'Shea}, {Oshino}, {Ossokine}, {Osthelder},
  {Otabe}, {Ottaway}, {Overmier}, {Pace}, {Pagano}, {Page}, {Pagliaroli},
  {Pai}, {Pai}, {Palamos}, {Palashov}, {Palomba}, {Pan}, {Panda}, {Pang},
  {Pang}, {Pankow}, {Pannarale}, {Pant}, {Paoletti}, {Paoli}, {Paolone},
  {Parisi}, {Park}, {Parker}, {Pascucci}, {Pasqualetti}, {Passaquieti},
  {Passuello}, {Patel}, {Patricelli}, {Payne}, {Pechsiri}, {Pedraza},
  {Pegoraro}, {Pele}, {Arellano}, {Penn}, {Perego}, {Pereira}, {Pereira},
  {Perez}, {P{\'e}rigois}, {Perreca}, {Perri{\`e}s}, {Petermann}, {Petterson},
  {Pfeiffer}, {Pham}, {Phukon}, {Piccinni}, {Pichot}, {Piendibene},
  {Piergiovanni}, {Pierini}, {Pierro}, {Pillant}, {Pilo}, {Pinard}, {Pinto},
  {Piotrzkowski}, {Piotrzkowski}, {Pirello}, {Pitkin}, {Placidi}, {Plastino},
  {Pluchar}, {Poggiani}, {Polini}, {Pong}, {Ponrathnam}, {Popolizio}, {Porter},
  {Powell}, {Pracchia}, {Pradier}, {Prajapati}, {Prasai}, {Prasanna},
  {Pratten}, {Prestegard}, {Principe}, {Prodi}, {Prokhorov}, {Prosposito},
  {Prudenzi}, {Puecher}, {Punturo}, {Puosi}, {Puppo}, {P{\"u}rrer}, {Qi},
  {Quetschke}, {Quinonez}, {Quitzow-James}, {Raab}, {Raaijmakers}, {Radkins},
  {Radulesco}, {Raffai}, {Rail}, {Raja}, {Rajan}, {Ramirez}, {Ramirez},
  {Ramos-Buades}, {Rana}, {Rapagnani}, {Rapol}, {Ratto}, {Raymond}, {Raza},
  {Razzano}, {Read}, {Rees}, {Regimbau}, {Rei}, {Reid}, {Reitze}, {Relton},
  {Rettegno}, {Ricci}, {Richardson}, {Richardson}, {Richardson}, {Ricker},
  {Riemenschneider}, {Riles}, {Rizzo}, {Robertson}, {Robie}, {Robinet},
  {Rocchi}, {Rocha}, {Rodriguez}, {Rodriguez-Soto}, {Rolland}, {Rollins},
  {Roma}, {Romanelli}, {Romano}, {Romano}, {Romel}, {Romero}, {Romero-Shaw},
  {Romie}, {Rose}, {Rosi{\'n}ska}, {Rosofsky}, {Ross}, {Rowan}, {Rowlinson},
  {Roy}, {Roy}, {Rozza}, {Ruggi}, {Ryan}, {Sachdev}, {Sadecki}, {Sadiq},
  {Sago}, {Saito}, {Saito}, {Sakai}, {Sakai}, {Sakellariadou}, {Sakuno},
  {Salafia}, {Salconi}, {Saleem}, {Salemi}, {Samajdar}, {Sanchez}, {Sanchez},
  {Sanchez}, {Sanchis-Gual}, {Sanders}, {Sanuy}, {Saravanan}, {Sarin},
  {Sassolas}, {Satari}, {Sato}, {Sato}, {Sauter}, {Savage}, {Savant}, {Sawada},
  {Sawant}, {Sawant}, {Sayah}, {Schaetzl}, {Scheel}, {Scheuer},
  {Schindler-Tyka}, {Schmidt}, {Schnabel}, {Schneewind}, {Schofield},
  {Sch{\"o}nbeck}, {Schulte}, {Schutz}, {Schwartz}, {Scott}, {Scott},
  {Seglar-Arroyo}, {Seidel}, {Sekiguchi}, {Sekiguchi}, {Sellers}, {Sengupta},
  {Sennett}, {Sentenac}, {Seo}, {Sequino}, {Sergeev}, {Setyawati}, {Shaffer},
  {Shahriar}, {Shams}, {Shao}, {Sharifi}, {Sharma}, {Sharma}, {Shawhan},
  {Shcheblanov}, {Shen}, {Shibagaki}, {Shikauchi}, {Shimizu}, {Shimoda},
  {Shimode}, {Shink}, {Shinkai}, {Shishido}, {Shoda}, {Shoemaker}, {Shoemaker},
  {Shukla}, {Shyamsundar}, {Sieniawska}, {Sigg}, {Singer}, {Singh}, {Singh},
  {Singha}, {Sintes}, {Sipala}, {Skliris}, {Slagmolen}, {Slaven-Blair},
  {Smetana}, {Smith}, {Smith}, {Somala}, {Somiya}, {Son}, {Soni}, {Soni},
  {Sorazu}, {Sordini}, {Sorrentino}, {Sorrentino}, {Sotani}, {Soulard},
  {Souradeep}, {Sowell}, {Spagnuolo}, {Spencer}, {Spera}, {Srivastava},
  {Srivastava}, {Staats}, {Stachie}, {Steer}, {Steinlechner}, {Steinlechner},
  {Stops}, {Stover}, {Strain}, {Strang}, {Stratta}, {Strunk}, {Sturani},
  {Stuver}, {S{\"u}dbeck}, {Sudhagar}, {Sudhir}, {Sugimoto}, {Suh},
  {Summerscales}, {Sun}, {Sun}, {Sunil}, {Sur}, {Suresh}, {Sutton}, {Suzuki},
  {Suzuki}, {Swinkels}, {Szczepa{\'n}czyk}, {Szewczyk}, {Tacca}, {Tagoshi},
  {Tait}, {Takahashi}, {Takahashi}, {Takamori}, {Takano}, {Takeda}, {Takeda},
  {Talbot}, {Tanaka}, {Tanaka}, {Tanaka}, {Tanaka}, {Tanaka}, {Tanasijczuk},
  {Tanioka}, {Tanner}, {Tao}, {Tapia}, {Martin}, {Martin}, {Tasson}, {Telada},
  {Tenorio}, {Terkowski}, {Test}, {Thirugnanasambandam}, {Thomas}, {Thomas},
  {Thompson}, {Thondapu}, {Thorne}, {Thrane}, {Tiwari}, {Tiwari}, {Tiwari},
  {Toland}, {Tolley}, {Tomaru}, {Tomigami}, {Tomura}, {Tonelli},
  {Torres-Forn{\'e}}, {Torrie}, {E Melo}, {T{\"o}yr{\"a}}, {Trapananti},
  {Travasso}, {Traylor}, {Tringali}, {Tripathee}, {Troiano}, {Trovato},
  {Trozzo}, {Trudeau}, {Tsai}, {Tsai}, {Tsang}, {Tsang}, {Tsao}, {Tse}, {Tso},
  {Tsubono}, {Tsuchida}, {Tsukada}, {Tsuna}, {Tsutsui}, {Tsuzuki}, {Turconi},
  {Tuyenbayev}, {Ubhi}, {Uchikata}, {Uchiyama}, {Udall}, {Ueda}, {Uehara},
  {Ueno}, {Ueshima}, {Ugolini}, {Unnikrishnan}, {Uraguchi}, {Urban}, {Ushiba},
  {Usman}, {Utina}, {Vahlbruch}, {Vajente}, {Vajpeyi}, {Valdes}, {Valentini},
  {Valsan}, {van Bakel}, {van Beuzekom}, {van den Brand}, {van den Broeck},
  {van Remortel}, {Vander-Hyde}, {van der Schaaf}, {van Heijningen}, {van
  Putten}, {Vardaro}, {Vargas}, {Varma}, {Vas{\'u}th}, {Vecchio}, {Vedovato},
  {Veitch}, {Veitch}, {Venkateswara}, {Venneberg}, {Venugopalan}, {Verkindt},
  {Verma}, {Veske}, {Vetrano}, {Vicer{\'e}}, {Viets}, {Villa-Ortega}, {Vinet},
  {Vitale}, {Vo}, {Vocca}, {von Reis}, {von Wrangel}, {Vorvick}, {Vyatchanin},
  {Wade}, {Wade}, {Wagner}, {Walet}, {Walker}, {Wallace}, {Wallace}, {Walsh},
  {Wang}, {Wang}, {Wang}, {Ward}, {Warner}, {Was}, {Washimi}, {Washington},
  {Watchi}, {Weaver}, {Wei}, {Weinert}, {Weinstein}, {Weiss}, {Weller},
  {Wellmann}, {Wen}, {We{\ss}els}, {Westhouse}, {Wette}, {Whelan}, {White},
  {Whiting}, {Whittle}, {Wilken}, {Williams}, {Williams}, {Williamson},
  {Willis}, {Willke}, {Wilson}, {Winkler}, {Wipf}, {Wlodarczyk}, {Woan},
  {Woehler}, {Wofford}, {Wong}, {Wu}, {Wu}, {Wu}, {Wu}, {Wysocki}, {Xiao},
  {Xu}, {Yamada}, {Yamamoto}, {Yamamoto}, {Yamamoto}, {Yamamoto}, {Yamashita},
  {Yamazaki}, {Yang}, {Yang}, {Yang}, {Yang}, {Yang}, {Yap}, {Yeeles},
  {Yelikar}, {Ying}, {Yokogawa}, {Yokoyama}, {Yokozawa}, {Yoon}, {Yoshioka},
  {Yu}, {Yu}, {Yuzurihara}, {Zadro{\.z}ny}, {Zanolin}, {Zeidler}, {Zelenova},
  {Zendri}, {Zevin}, {Zhan}, {Zhang}, {Zhang}, {Zhang}, {Zhang}, {Zhang},
  {Zhao}, {Zhao}, {Zhao}, {Zhao}, {Zhou}, {Zhu}, {Zhu}, {Zucker}, {Zweizig},
  {Ligo Scientific Collaboration}, {VIRGO Collaboration}, \& {Kagra
  Collaboration}}]{2021PhRvD.104b2004A}
{Abbott}, R., {Abbott}, T.~D., {Abraham}, S., {et~al.} 2021,
  \bibinfo{title}{{Upper limits on the isotropic gravitational-wave background
  from Advanced LIGO and Advanced Virgo's third observing run},} \prd, 104,
  022004, \dodoi{10.1103/PhysRevD.104.022004}

\bibitem[{A. {Afzal} {et~al.}(2023){Afzal}, {Agazie}, {Anumarlapudi},
  {Archibald}, {Arzoumanian}, {Baker}, {B{\'e}csy}, {Blanco-Pillado}, {Blecha},
  {Boddy}, {Brazier}, {Brook}, {Burke-Spolaor}, {Burnette}, {Case}, {Charisi},
  {Chatterjee}, {Chatziioannou}, {Cheeseboro}, {Chen}, {Cohen}, {Cordes},
  {Cornish}, {Crawford}, {Cromartie}, {Crowter}, {Cutler}, {Decesar}, {Degan},
  {Demorest}, {Deng}, {Dolch}, {Drachler}, {von Eckardstein}, {Ferrara},
  {Fiore}, {Fonseca}, {Freedman}, {Garver-Daniels}, {Gentile}, {Gersbach},
  {Glaser}, {Good}, {Guertin}, {G{\"u}ltekin}, {Hazboun}, {Hourihane}, {Islo},
  {Jennings}, {Johnson}, {Jones}, {Kaiser}, {Kaplan}, {Kelley}, {Kerr}, {Key},
  {Laal}, {Lam}, {Lamb}, {Lazio}, {Lee}, {Lewandowska}, {Lino Dos Santos},
  {Littenberg}, {Liu}, {Lorimer}, {Luo}, {Lynch}, {Ma}, {Madison}, {McEwen},
  {McKee}, {McLaughlin}, {McMann}, {Meyers}, {Meyers}, {Mingarelli},
  {Mitridate}, {Nay}, {Natarajan}, {Ng}, {Nice}, {Ocker}, {Olum}, {Pennucci},
  {Perera}, {Petrov}, {Pol}, {Radovan}, {Ransom}, {Ray}, {Romano}, {Sardesai},
  {Schmiedekamp}, {Schmiedekamp}, {Schmitz}, {Schr{\"o}der}, {Schult},
  {Shapiro-Albert}, {Siemens}, {Simon}, {Siwek}, {Stairs}, {Stinebring},
  {Stovall}, {Stratmann}, {Sun}, {Susobhanan}, {Swiggum}, {Taylor}, {Taylor},
  {Trickle}, {Turner}, {Unal}, {Vallisneri}, {Verma}, {Vigeland}, {Wahl},
  {Wang}, {Witt}, {Wright}, {Young}, {Zurek}, \& {Nanograv
  Collaboration}}]{2023ApJ...951L..11A}
{Afzal}, A., {Agazie}, G., {Anumarlapudi}, A., {et~al.} 2023,
  \bibinfo{title}{{The NANOGrav 15 yr Data Set: Search for Signals from New
  Physics},} \apjl, 951, L11, \dodoi{10.3847/2041-8213/acdc91}

\bibitem[{G. {Agazie} {et~al.}(2023){Agazie}, {Anumarlapudi}, {Archibald},
  {Arzoumanian}, {Baker}, {B{\'e}csy}, {Blecha}, {Brazier}, {Brook},
  {Burke-Spolaor}, {Burnette}, {Case}, {Charisi}, {Chatterjee},
  {Chatziioannou}, {Cheeseboro}, {Chen}, {Cohen}, {Cordes}, {Cornish},
  {Crawford}, {Cromartie}, {Crowter}, {Cutler}, {Decesar}, {Degan}, {Demorest},
  {Deng}, {Dolch}, {Drachler}, {Ellis}, {Ferrara}, {Fiore}, {Fonseca},
  {Freedman}, {Garver-Daniels}, {Gentile}, {Gersbach}, {Glaser}, {Good},
  {G{\"u}ltekin}, {Hazboun}, {Hourihane}, {Islo}, {Jennings}, {Johnson},
  {Jones}, {Kaiser}, {Kaplan}, {Kelley}, {Kerr}, {Key}, {Klein}, {Laal}, {Lam},
  {Lamb}, {Lazio}, {Lewandowska}, {Littenberg}, {Liu}, {Lommen}, {Lorimer},
  {Luo}, {Lynch}, {Ma}, {Madison}, {Mattson}, {McEwen}, {McKee}, {McLaughlin},
  {McMann}, {Meyers}, {Meyers}, {Mingarelli}, {Mitridate}, {Natarajan}, {Ng},
  {Nice}, {Ocker}, {Olum}, {Pennucci}, {Perera}, {Petrov}, {Pol}, {Radovan},
  {Ransom}, {Ray}, {Romano}, {Sardesai}, {Schmiedekamp}, {Schmiedekamp},
  {Schmitz}, {Schult}, {Shapiro-Albert}, {Siemens}, {Simon}, {Siwek}, {Stairs},
  {Stinebring}, {Stovall}, {Sun}, {Susobhanan}, {Swiggum}, {Taylor}, {Taylor},
  {Turner}, {Unal}, {Vallisneri}, {van Haasteren}, {Vigeland}, {Wahl}, {Wang},
  {Witt}, {Young}, \& {Nanograv Collaboration}}]{2023ApJ...951L...8A}
{Agazie}, G., {Anumarlapudi}, A., {Archibald}, A.~M., {et~al.} 2023,
  \bibinfo{title}{{The NANOGrav 15 yr Data Set: Evidence for a
  Gravitational-wave Background},} \apjl, 951, L8,
  \dodoi{10.3847/2041-8213/acdac6}

\bibitem[{G. {Agazie} {et~al.}(2024){Agazie}, {Antoniadis}, {Anumarlapudi},
  {Archibald}, {Arumugam}, {Arumugam}, {Arzoumanian}, {Askew}, {Babak},
  {Bagchi}, {Bailes}, {Bak Nielsen}, {Baker}, {Bassa}, {Bathula}, {B{\'e}csy},
  {Berthereau}, {Bhat}, {Blecha}, {Bonetti}, {Bortolas}, {Brazier}, {Brook},
  {Burgay}, {Burke-Spolaor}, {Burnette}, {Caballero}, {Cameron}, {Case},
  {Chalumeau}, {Champion}, {Chanlaridis}, {Charisi}, {Chatterjee},
  {Chatziioannou}, {Cheeseboro}, {Chen}, {Chen}, {Cognard}, {Cohen}, {Coles},
  {Cordes}, {Cornish}, {Crawford}, {Cromartie}, {Crowter}, {Cury{\l}o},
  {Cutler}, {Dai}, {Dandapat}, {Deb}, {DeCesar}, {DeGan}, {Demorest}, {Deng},
  {Desai}, {Desvignes}, {Dey}, {Dhanda-Batra}, {Di Marco}, {Dolch}, {Drachler},
  {Dwivedi}, {Ellis}, {Falxa}, {Feng}, {Ferdman}, {Ferrara}, {Fiore},
  {Fonseca}, {Franchini}, {Freedman}, {Gair}, {Garver-Daniels}, {Gentile},
  {Gersbach}, {Glaser}, {Good}, {Goncharov}, {Gopakumar}, {Graikou},
  {Griessmeier}, {Guillemot}, {G{\"u}ltekin}, {Guo}, {Gupta}, {Grunthal},
  {Hazboun}, {Hisano}, {Hobbs}, {Hourihane}, {Hu}, {Iraci}, {Islo},
  {Izquierdo-Villalba}, {Jang}, {Jawor}, {Janssen}, {Jennings}, {Jessner},
  {Johnson}, {Jones}, {Joshi}, {Kaiser}, {Kaplan}, {Kapur}, {Kareem},
  {Karuppusamy}, {Keane}, {Keith}, {Kelley}, {Kerr}, {Key}, {Kharbanda},
  {Kikunaga}, {Klein}, {Kolhe}, {Kramer}, {Krishnakumar}, {Kulkarni}, {Laal},
  {Lackeos}, {Lam}, {Lamb}, {Larsen}, {Lazio}, {Lee}, {Levin}, {Lewandowska},
  {Littenberg}, {Liu}, {Liu}, {Liu}, {Lommen}, {Lorimer}, {Lower}, {Luo},
  {Luo}, {Lynch}, {Lyne}, {Ma}, {Maan}, {Madison}, {Main}, {Manchester},
  {Mandow}, {Mattson}, {McEwen}, {McKee}, {McLaughlin}, {McMann}, {Meyers},
  {Meyers}, {Mickaliger}, {Miles}, {Mingarelli}, {Mitridate}, {Natarajan},
  {Nathan}, {Ng}, {Nice}, {Ni{\c{t}}u}, {Nobleson}, {Ocker}, {Olum},
  {Os{\l}owski}, {Paladi}, {Parthasarathy}, {Pennucci}, {Perera}, {Perrodin},
  {Petiteau}, {Petrov}, {Pol}, {Porayko}, {Possenti}, {Prabu}, {Quelquejay
  Leclere}, {Radovan}, {Rana}, {Ransom}, {Ray}, {Reardon}, {Rogers}, {Romano},
  {Russell}, {Samajdar}, {Sanidas}, {Sardesai}, {Schmiedekamp}, {Schmiedekamp},
  {Schmitz}, {Schult}, {Sesana}, {Shaifullah}, {Shannon}, {Shapiro-Albert},
  {Siemens}, {Simon}, {Singha}, {Siwek}, {Speri}, {Spiewak}, {Srivastava},
  {Stairs}, {Stappers}, {Stinebring}, {Stovall}, {Sun}, {Surnis}, {Susarla},
  {Susobhanan}, {Swiggum}, {Takahashi}, {Tarafdar}, {Taylor}, {Taylor},
  {Theureau}, {Thrane}, {Thyagarajan}, {Tiburzi}, {Toomey}, {Turner}, {Unal},
  {Vallisneri}, {van der Wateren}, {van Haasteren}, {Vecchio}, {Venkatraman
  Krishnan}, {Verbiest}, {Vigeland}, {Wahl}, {Wang}, {Wang}, {Witt}, {Wang},
  {Wang}, {Wayt}, {Wu}, {Young}, {Zhang}, {Zhang}, {Zhu}, {Zic}, \&
  {International Pulsar Timing Array Collaboration}}]{2024ApJ...966..105A}
{Agazie}, G., {Antoniadis}, J., {Anumarlapudi}, A., {et~al.} 2024,
  \bibinfo{title}{{Comparing Recent Pulsar Timing Array Results on the
  Nanohertz Stochastic Gravitational-wave Background},} \apj, 966, 105,
  \dodoi{10.3847/1538-4357/ad36be}

\bibitem[{K.~N. {Ananda} {et~al.}(2007){Ananda}, {Clarkson}, \&
  {Wands}}]{2007PhRvD..75l3518A}
{Ananda}, K.~N., {Clarkson}, C., \& {Wands}, D. 2007,
  \bibinfo{title}{{Cosmological gravitational wave background from primordial
  density perturbations},} \prd, 75, 123518, \dodoi{10.1103/PhysRevD.75.123518}

\bibitem[{J. {Antoniadis} {et~al.}(2023){Antoniadis}, {Arumugam}, {Arumugam},
  {Babak}, {Bagchi}, {Bak Nielsen}, {Bassa}, {Bathula}, {Berthereau},
  {Bonetti}, {Bortolas}, {Brook}, {Burgay}, {Caballero}, {Chalumeau},
  {Champion}, {Chanlaridis}, {Chen}, {Cognard}, {Dandapat}, {Deb}, {Desai},
  {Desvignes}, {Dhanda-Batra}, {Dwivedi}, {Falxa}, {Ferdman}, {Franchini},
  {Gair}, {Goncharov}, {Gopakumar}, {Graikou}, {Grie{\ss}meier}, {Guillemot},
  {Guo}, {Gupta}, {Hisano}, {Hu}, {Iraci}, {Izquierdo-Villalba}, {Jang},
  {Jawor}, {Janssen}, {Jessner}, {Joshi}, {Kareem}, {Karuppusamy}, {Keane},
  {Keith}, {Kharbanda}, {Kikunaga}, {Kolhe}, {Kramer}, {Krishnakumar},
  {Lackeos}, {Lee}, {Liu}, {Liu}, {Lyne}, {McKee}, {Maan}, {Main},
  {Mickaliger}, {Ni{\c{t}}u}, {Nobleson}, {Paladi}, {Parthasarathy}, {Perera},
  {Perrodin}, {Petiteau}, {Porayko}, {Possenti}, {Prabu}, {Quelquejay Leclere},
  {Rana}, {Samajdar}, {Sanidas}, {Sesana}, {Shaifullah}, {Singha}, {Speri},
  {Spiewak}, {Srivastava}, {Stappers}, {Surnis}, {Susarla}, {Susobhanan},
  {Takahashi}, {Tarafdar}, {Theureau}, {Tiburzi}, {van der Wateren}, {Vecchio},
  {Venkatraman Krishnan}, {Verbiest}, {Wang}, {Wang}, \& {Wu}}]{2023EPTA_DR2}
{Antoniadis}, J., {Arumugam}, P., {Arumugam}, S., {et~al.} 2023,
  \bibinfo{title}{{The second data release from the European Pulsar Timing
  Array III. Search for gravitational wave signals},}, 2.0 Zenodo,
  \dodoi{10.5281/zenodo.8091568}

\bibitem[{Z. {Arzoumanian} {et~al.}(2021){Arzoumanian}, {Baker}, {Blumer},
  {B{\'e}csy}, {Brazier}, {Brook}, {Burke-Spolaor}, {Charisi}, {Chatterjee},
  {Chen}, {Cordes}, {Cornish}, {Crawford}, {Cromartie}, {Decesar}, {Demorest},
  {Dolch}, {Ellis}, {Ferrara}, {Fiore}, {Fonseca}, {Garver-Daniels}, {Gentile},
  {Good}, {Hazboun}, {Holgado}, {Islo}, {Jennings}, {Jones}, {Kaiser},
  {Kaplan}, {Kelley}, {Key}, {Laal}, {Lam}, {Lazio}, {Lee}, {Lorimer}, {Luo},
  {Lynch}, {Madison}, {McLaughlin}, {Mingarelli}, {Mitridate}, {Ng}, {Nice},
  {Pennucci}, {Pol}, {Ransom}, {Ray}, {Shapiro-Albert}, {Siemens}, {Simon},
  {Spiewak}, {Stairs}, {Stinebring}, {Stovall}, {Sun}, {Swiggum}, {Taylor},
  {Turner}, {Vallisneri}, {Vigeland}, {Witt}, {Zurek}, \& {Nanograv
  Collaboration}}]{2021PhRvL.127y1302A}
{Arzoumanian}, Z., {Baker}, P.~T., {Blumer}, H., {et~al.} 2021,
  \bibinfo{title}{{Searching for Gravitational Waves from Cosmological Phase
  Transitions with the NANOGrav 12.5-Year Dataset},} \prl, 127, 251302,
  \dodoi{10.1103/PhysRevLett.127.251302}

\bibitem[{P. {Auclair} {et~al.}(2020){Auclair}, {Blanco-Pillado}, {Figueroa},
  {Jenkins}, {Lewicki}, {Sakellariadou}, {Sanidas}, {Sousa}, {Steer},
  {Wachter}, {Kuroyanagi}, \& {LISA Cosmology Working
  Group}}]{2020JCAP...04..034A}
{Auclair}, P., {Blanco-Pillado}, J.~J., {Figueroa}, D.~G., {et~al.} 2020,
  \bibinfo{title}{{Probing the gravitational wave background from cosmic
  strings with LISA},} \jcap, 2020, 034, \dodoi{10.1088/1475-7516/2020/04/034}

\bibitem[{P. {Auclair} {et~al.}(2023){Auclair}, {Bacon}, {Baker}, {Barreiro},
  {Bartolo}, {Belgacem}, {Bellomo}, {Ben-Dayan}, {Bertacca}, {Besancon},
  {Blanco-Pillado}, {Blas}, {Boileau}, {Calcagni}, {Caldwell}, {Caprini},
  {Carbone}, {Chang}, {Chen}, {Christensen}, {Clesse}, {Comelli}, {Congedo},
  {Contaldi}, {Crisostomi}, {Croon}, {Cui}, {Cusin}, {Cutting}, {Dalang}, {De
  Luca}, {Pozzo}, {Desjacques}, {Dimastrogiovanni}, {Dorsch}, {Ezquiaga},
  {Fasiello}, {Figueroa}, {Flauger}, {Franciolini}, {Frusciante}, {Fumagalli},
  {Garc{\'\i}a-Bellido}, {Gould}, {Holz}, {Iacconi}, {Jain}, {Jenkins},
  {Jinno}, {Joana}, {Karnesis}, {Konstandin}, {Koyama}, {Kozaczuk},
  {Kuroyanagi}, {Laghi}, {Lewicki}, {Lombriser}, {Madge}, {Maggiore},
  {Malhotra}, {Mancarella}, {Mandic}, {Mangiagli}, {Matarrese}, {Mazumdar},
  {Mukherjee}, {Musco}, {Nardini}, {No}, {Papanikolaou}, {Peloso}, {Pieroni},
  {Pilo}, {Raccanelli}, {Renaux-Petel}, {Renzini}, {Ricciardone}, {Riotto},
  {Romano}, {Rollo}, {Pol}, {Morales}, {Sakellariadou}, {Saltas}, {Scalisi},
  {Schmitz}, {Schwaller}, {Sergijenko}, {Servant}, {Simakachorn}, {Sorbo},
  {Sousa}, {Speri}, {Steer}, {Tamanini}, {Tasinato}, {Torrado}, {Unal},
  {Vennin}, {Vernieri}, {Vernizzi}, {Volonteri}, {Wachter}, {Wands},
  {Witkowski}, {Zumalac{\'a}rregui}, {Annis}, {Ares}, {Avelino}, {Avgoustidis},
  {Barausse}, {Bonilla}, {Bonvin}, {Bosso}, {Calabrese},
  {{\c{C}}al{\i}{\c{s}}kan}, {Cembranos}, {Chala}, {Chernoff}, {Clough},
  {Criswell}, {Das}, {Silva}, {Dayal}, {Domcke}, {Durrer}, {Easther},
  {Escoffier}, {Ferrans}, {Fryer}, {Gair}, {Gordon}, {Hendry}, {Hindmarsh},
  {Hooper}, {Kajfasz}, {Kopp}, {Koushiappas}, {Kumar}, {Kunz}, {Lagos},
  {Lilley}, {Lizarraga}, {Lobo}, {Maleknejad}, {Martins}, {Meerburg}, {Meyer},
  {Mimoso}, {Nesseris}, {Nunes}, {Oikonomou}, {Orlando}, {{\"O}zsoy},
  {Pacucci}, {Palmese}, {Petiteau}, {Pinol}, {Zwart}, {Pratten}, {Prokopec},
  {Quenby}, {Rastgoo}, {Roest}, {Rummukainen}, {Schimd}, {Secroun}, {Sesana},
  {Sopuerta}, {Tereno}, {Tolley}, {Urrestilla}, {Vagenas}, {van de Vis}, {van
  de Weygaert}, {Wardell}, {Weir}, {White}, {{\'S}wie{\.z}ewska}, {Zhdanov}, \&
  {The LISA Cosmology Working Group}}]{2023LRR....26....5A}
{Auclair}, P., {Bacon}, D., {Baker}, T., {et~al.} 2023,
  \bibinfo{title}{{Cosmology with the Laser Interferometer Space Antenna},}
  Living Reviews in Relativity, 26, 5, \dodoi{10.1007/s41114-023-00045-2}

\bibitem[{D. {Baumann} {et~al.}(2007){Baumann}, {Steinhardt}, {Takahashi}, \&
  {Ichiki}}]{2007PhRvD..76h4019B}
{Baumann}, D., {Steinhardt}, P., {Takahashi}, K., \& {Ichiki}, K. 2007,
  \bibinfo{title}{{Gravitational wave spectrum induced by primordial scalar
  perturbations},} \prd, 76, 084019, \dodoi{10.1103/PhysRevD.76.084019}

\bibitem[{M. {Baumgart} {et~al.}(2022){Baumgart}, {Heckman}, \&
  {Thomas}}]{2022JCAP...07..034B}
{Baumgart}, M., {Heckman}, J.~J., \& {Thomas}, L. 2022, \bibinfo{title}{{CFTs
  blueshift tensor fluctuations universally},} \jcap, 2022, 034,
  \dodoi{10.1088/1475-7516/2022/07/034}

\bibitem[{M. {Benetti} {et~al.}(2022){Benetti}, {Graef}, \&
  {Vagnozzi}}]{2022PhRvD.105d3520B}
{Benetti}, M., {Graef}, L.~L., \& {Vagnozzi}, S. 2022,
  \bibinfo{title}{{Primordial gravitational waves from NANOGrav: A broken
  power-law approach},} \prd, 105, 043520, \dodoi{10.1103/PhysRevD.105.043520}

\bibitem[{J.~J. {Bennett} {et~al.}(2021){Bennett}, {Buldgen}, {de Salas},
  {Drewes}, {Gariazzo}, {Pastor}, \& {Wong}}]{2021JCAP...04..073B}
{Bennett}, J.~J., {Buldgen}, G., {de Salas}, P.~F., {et~al.} 2021,
  \bibinfo{title}{{Towards a precision calculation of the effective number of
  neutrinos N$_{eff}$ in the Standard Model. Part II. Neutrino decoupling in
  the presence of flavour oscillations and finite-temperature QED},} \jcap,
  2021, 073, \dodoi{10.1088/1475-7516/2021/04/073}

\bibitem[{S. {Blasi} {et~al.}(2021){Blasi}, {Brdar}, \&
  {Schmitz}}]{2021PhRvL.126d1305B}
{Blasi}, S., {Brdar}, V., \& {Schmitz}, K. 2021, \bibinfo{title}{{Has NANOGrav
  Found First Evidence for Cosmic Strings?},} \prl, 126, 041305,
  \dodoi{10.1103/PhysRevLett.126.041305}

\bibitem[{L.~A. {Boyle} \& A. {Buonanno}(2008){Boyle} \&
  {Buonanno}}]{2008PhRvD..78d3531B}
{Boyle}, L.~A., \& {Buonanno}, A. 2008, \bibinfo{title}{{Relating gravitational
  wave constraints from primordial nucleosynthesis, pulsar timing, laser
  interferometers, and the CMB: Implications for the early universe},} \prd,
  78, 043531, \dodoi{10.1103/PhysRevD.78.043531}

\bibitem[{L.~A. {Boyle} \& P.~J. {Steinhardt}(2008){Boyle} \&
  {Steinhardt}}]{2008PhRvD..77f3504B}
{Boyle}, L.~A., \& {Steinhardt}, P.~J. 2008, \bibinfo{title}{{Probing the early
  universe with inflationary gravitational waves},} \prd, 77, 063504,
  \dodoi{10.1103/PhysRevD.77.063504}

\bibitem[{L.~A. {Boyle} {et~al.}(2004){Boyle}, {Steinhardt}, \&
  {Turok}}]{2004PhRvD..69l7302B}
{Boyle}, L.~A., {Steinhardt}, P.~J., \& {Turok}, N. 2004,
  \bibinfo{title}{{Cosmic gravitational-wave background in a cyclic universe},}
  \prd, 69, 127302, \dodoi{10.1103/PhysRevD.69.127302}

\bibitem[{R.-G. {Cai} {et~al.}(2019){Cai}, {Pi}, \&
  {Sasaki}}]{2019PhRvL.122t1101C}
{Cai}, R.-G., {Pi}, S., \& {Sasaki}, M. 2019, \bibinfo{title}{{Gravitational
  Waves Induced by Non-Gaussian Scalar Perturbations},} \prl, 122, 201101,
  \dodoi{10.1103/PhysRevLett.122.201101}

\bibitem[{Y.-F. {Cai} {et~al.}(2015){Cai}, {Gong}, {Pi}, {Saridakis}, \&
  {Wu}}]{2015NuPhB.900..517C}
{Cai}, Y.-F., {Gong}, J.-O., {Pi}, S., {Saridakis}, E.~N., \& {Wu}, S.-Y. 2015,
  \bibinfo{title}{{On the possibility of blue tensor spectrum within single
  field inflation},} Nuclear Physics B, 900, 517,
  \dodoi{10.1016/j.nuclphysb.2015.09.025}

\bibitem[{Y.-F. {Cai} {et~al.}(2023){Cai}, {He}, {Ma}, {Yan}, \&
  {Yuan}}]{2023SciBu..68.2929Y}
{Cai}, Y.-F., {He}, X.-C., {Ma}, X.-H., {Yan}, S.-F., \& {Yuan}, G.-W. 2023,
  \bibinfo{title}{{Limits on scalar-induced gravitational waves from the
  stochastic background by pulsar timing array observations},} Science
  Bulletin, 68, 2929, \dodoi{10.1016/j.scib.2023.10.027}

\bibitem[{R.~R. {Caldwell} \& C. {Devulder}(2018){Caldwell} \&
  {Devulder}}]{2018PhRvD..97b3532C}
{Caldwell}, R.~R., \& {Devulder}, C. 2018, \bibinfo{title}{{Axion gauge field
  inflation and gravitational leptogenesis: A lower bound on B modes from the
  matter-antimatter asymmetry of the Universe},} \prd, 97, 023532,
  \dodoi{10.1103/PhysRevD.97.023532}

\bibitem[{C. {Caprini} \& D.~G. {Figueroa}(2018){Caprini} \&
  {Figueroa}}]{2018CQGra..35p3001C}
{Caprini}, C., \& {Figueroa}, D.~G. 2018, \bibinfo{title}{{Cosmological
  backgrounds of gravitational waves},} Classical and Quantum Gravity, 35,
  163001, \dodoi{10.1088/1361-6382/aac608}

\bibitem[{R.~T. {Co} {et~al.}(2022){Co}, {Dunsky}, {Fernandez}, {Ghalsasi},
  {Hall}, {Harigaya}, \& {Shelton}}]{2022JHEP...09..116C}
{Co}, R.~T., {Dunsky}, D., {Fernandez}, N., {et~al.} 2022,
  \bibinfo{title}{{Gravitational wave and CMB probes of axion kination},}
  Journal of High Energy Physics, 2022, 116, \dodoi{10.1007/JHEP09(2022)116}

\bibitem[{R.~T. {Co} {et~al.}(2020){Co}, {Hall}, \&
  {Harigaya}}]{2020PhRvL.124y1802C}
{Co}, R.~T., {Hall}, L.~J., \& {Harigaya}, K. 2020, \bibinfo{title}{{Axion
  Kinetic Misalignment Mechanism},} \prl, 124, 251802,
  \dodoi{10.1103/PhysRevLett.124.251802}

\bibitem[{J.~L. {Cook} \& L. {Sorbo}(2012){Cook} \&
  {Sorbo}}]{2012PhRvD..85b3534C}
{Cook}, J.~L., \& {Sorbo}, L. 2012, \bibinfo{title}{{Particle production during
  inflation and gravitational waves detectable by ground-based
  interferometers},} \prd, 85, 023534, \dodoi{10.1103/PhysRevD.85.023534}

\bibitem[{P. {Creminelli} {et~al.}(2014){Creminelli}, {Gleyzes}, {Nore{\~n}a},
  \& {Vernizzi}}]{2014PhRvL.113w1301C}
{Creminelli}, P., {Gleyzes}, J., {Nore{\~n}a}, J., \& {Vernizzi}, F. 2014,
  \bibinfo{title}{{Resilience of the Standard Predictions for Primordial Tensor
  Modes},} \prl, 113, 231301, \dodoi{10.1103/PhysRevLett.113.231301}

\bibitem[{Y. {Cui} {et~al.}(2018){Cui}, {Lewicki}, {Morrissey}, \&
  {Wells}}]{2018PhRvD..97l3505C}
{Cui}, Y., {Lewicki}, M., {Morrissey}, D.~E., \& {Wells}, J.~D. 2018,
  \bibinfo{title}{{Cosmic archaeology with gravitational waves from cosmic
  strings},} \prd, 97, 123505, \dodoi{10.1103/PhysRevD.97.123505}

\bibitem[{ {DESI Collaboration} {et~al.}(2024){DESI Collaboration}, {Adame},
  {Aguilar}, {Ahlen}, {Alam}, {Alexander}, {Allende Prieto}, {Alvarez},
  {Alves}, {Anand}, \& et~al.}]{2024arXiv241112022D}
{DESI Collaboration}, {Adame}, A.~G., {Aguilar}, J., {et~al.} 2024,
  \bibinfo{title}{{DESI 2024 VII: Cosmological Constraints from the Full-Shape
  Modeling of Clustering Measurements},} arXiv e-prints, arXiv:2411.12022,
  \dodoi{10.48550/arXiv.2411.12022}

\bibitem[{G. {Dom{\`e}nech}(2021){Dom{\`e}nech}}]{2021Univ....7..398D}
{Dom{\`e}nech}, G. 2021, \bibinfo{title}{{Scalar Induced Gravitational Waves
  Review},} Universe, 7, 398, \dodoi{10.3390/universe7110398}

\bibitem[{S. {Dutta} \& R.~J. {Scherrer}(2010){Dutta} \&
  {Scherrer}}]{2010PhRvD..82h3501D}
{Dutta}, S., \& {Scherrer}, R.~J. 2010, \bibinfo{title}{{Big bang
  nucleosynthesis with a stiff fluid},} \prd, 82, 083501,
  \dodoi{10.1103/PhysRevD.82.083501}

\bibitem[{R. {Easther} {et~al.}(2007){Easther}, {Giblin}, \&
  {Lim}}]{2007PhRvL..99v1301E}
{Easther}, R., {Giblin}, Jr., J.~T., \& {Lim}, E.~A. 2007,
  \bibinfo{title}{{Gravitational Wave Production at the End of Inflation},}
  \prl, 99, 221301, \dodoi{10.1103/PhysRevLett.99.221301}

\bibitem[{J. {Ellis} \& M. {Lewicki}(2021){Ellis} \&
  {Lewicki}}]{2021PhRvL.126d1304E}
{Ellis}, J., \& {Lewicki}, M. 2021, \bibinfo{title}{{Cosmic String
  Interpretation of NANOGrav Pulsar Timing Data},} \prl, 126, 041304,
  \dodoi{10.1103/PhysRevLett.126.041304}

\bibitem[{J. {Ellis} {et~al.}(2024){Ellis}, {Fairbairn}, {Franciolini},
  {H{\"u}tsi}, {Iovino}, {Lewicki}, {Raidal}, {Urrutia}, {Vaskonen}, \&
  {Veerm{\"a}e}}]{2024PhRvD.109b3522E}
{Ellis}, J., {Fairbairn}, M., {Franciolini}, G., {et~al.} 2024,
  \bibinfo{title}{{What is the source of the PTA GW signal?},} \prd, 109,
  023522, \dodoi{10.1103/PhysRevD.109.023522}

\bibitem[{ {EPTA Collaboration} {et~al.}(2023){EPTA Collaboration}, {InPTA
  Collaboration}, {Antoniadis}, {Arumugam}, {Arumugam}, {Babak}, {Bagchi}, {Bak
  Nielsen}, {Bassa}, {Bathula}, {Berthereau}, {Bonetti}, {Bortolas}, {Brook},
  {Burgay}, {Caballero}, {Chalumeau}, {Champion}, {Chanlaridis}, {Chen},
  {Cognard}, {Dandapat}, {Deb}, {Desai}, {Desvignes}, {Dhanda-Batra},
  {Dwivedi}, {Falxa}, {Ferdman}, {Franchini}, {Gair}, {Goncharov}, {Gopakumar},
  {Graikou}, {Grie{\ss}meier}, {Guillemot}, {Guo}, {Gupta}, {Hisano}, {Hu},
  {Iraci}, {Izquierdo-Villalba}, {Jang}, {Jawor}, {Janssen}, {Jessner},
  {Joshi}, {Kareem}, {Karuppusamy}, {Keane}, {Keith}, {Kharbanda}, {Kikunaga},
  {Kolhe}, {Kramer}, {Krishnakumar}, {Lackeos}, {Lee}, {Liu}, {Liu}, {Lyne},
  {McKee}, {Maan}, {Main}, {Mickaliger}, {Ni{\c{t}}u}, {Nobleson}, {Paladi},
  {Parthasarathy}, {Perera}, {Perrodin}, {Petiteau}, {Porayko}, {Possenti},
  {Prabu}, {Quelquejay Leclere}, {Rana}, {Samajdar}, {Sanidas}, {Sesana},
  {Shaifullah}, {Singha}, {Speri}, {Spiewak}, {Srivastava}, {Stappers},
  {Surnis}, {Susarla}, {Susobhanan}, {Takahashi}, {Tarafdar}, {Theureau},
  {Tiburzi}, {van der Wateren}, {Vecchio}, {Venkatraman Krishnan}, {Verbiest},
  {Wang}, {Wang}, \& {Wu}}]{2023A&A...678A..50E}
{EPTA Collaboration}, {InPTA Collaboration}, {Antoniadis}, J., {et~al.} 2023,
  \bibinfo{title}{{The second data release from the European Pulsar Timing
  Array. III. Search for gravitational wave signals},} \aap, 678, A50,
  \dodoi{10.1051/0004-6361/202346844}

\bibitem[{ {EPTA Collaboration} {et~al.}(2024){EPTA Collaboration}, {InPTA
  Collaboration}, {Antoniadis}, {Arumugam}, {Arumugam}, {Babak}, {Bagchi}, {Bak
  Nielsen}, {Bassa}, {Bathula}, {Berthereau}, {Bonetti}, {Bortolas}, {Brook},
  {Burgay}, {Caballero}, {Chalumeau}, {Champion}, {Chanlaridis}, {Chen},
  {Cognard}, {Dandapat}, {Deb}, {Desai}, {Desvignes}, {Dhanda-Batra},
  {Dwivedi}, {Falxa}, {Ferdman}, {Franchini}, {Gair}, {Goncharov}, {Gopakumar},
  {Graikou}, {Grie{\ss}meier}, {Gualandris}, {Guillemot}, {Guo}, {Gupta},
  {Hisano}, {Hu}, {Iraci}, {Izquierdo-Villalba}, {Jang}, {Jawor}, {Janssen},
  {Jessner}, {Joshi}, {Kareem}, {Karuppusamy}, {Keane}, {Keith}, {Kharbanda},
  {Kikunaga}, {Kolhe}, {Kramer}, {Krishnakumar}, {Lackeos}, {Lee}, {Liu},
  {Liu}, {Lyne}, {McKee}, {Maan}, {Main}, {Mickaliger}, {Ni{\c{t}}u},
  {Nobleson}, {Paladi}, {Parthasarathy}, {Perera}, {Perrodin}, {Petiteau},
  {Porayko}, {Possenti}, {Prabu}, {Quelquejay Leclere}, {Rana}, {Samajdar},
  {Sanidas}, {Sesana}, {Shaifullah}, {Singha}, {Speri}, {Spiewak},
  {Srivastava}, {Stappers}, {Surnis}, {Susarla}, {Susobhanan}, {Takahashi},
  {Tarafdar}, {Theureau}, {Tiburzi}, {van der Wateren}, {Vecchio}, {Venkatraman
  Krishnan}, {Verbiest}, {Wang}, {Wang}, {Wu}, {Auclair}, {Barausse},
  {Caprini}, {Crisostomi}, {Fastidio}, {Khizriev}, {Middleton}, {Neronov},
  {Postnov}, {Roper Pol}, {Semikoz}, {Smarra}, {Steer}, {Truant}, \&
  {Valtolina}}]{2024A&A...685A..94E}
{EPTA Collaboration}, {InPTA Collaboration}, {Antoniadis}, J., {et~al.} 2024,
  \bibinfo{title}{{The second data release from the European Pulsar Timing
  Array. IV. Implications for massive black holes, dark matter, and the early
  Universe},} \aap, 685, A94, \dodoi{10.1051/0004-6361/202347433}

\bibitem[{D.~G. {Figueroa} {et~al.}(2024){Figueroa}, {Pieroni}, {Ricciardone},
  \& {Simakachorn}}]{2024PhRvL.132q1002F}
{Figueroa}, D.~G., {Pieroni}, M., {Ricciardone}, A., \& {Simakachorn}, P. 2024,
  \bibinfo{title}{{Cosmological Background Interpretation of Pulsar Timing
  Array Data},} \prl, 132, 171002, \dodoi{10.1103/PhysRevLett.132.171002}

\bibitem[{D.~G. {Figueroa} \& E.~H. {Tanin}(2019){Figueroa} \&
  {Tanin}}]{2019JCAP...08..011F}
{Figueroa}, D.~G., \& {Tanin}, E.~H. 2019, \bibinfo{title}{{Ability of LIGO and
  LISA to probe the equation of state of the early Universe},} \jcap, 2019,
  011, \dodoi{10.1088/1475-7516/2019/08/011}

\bibitem[{G. {Franciolini} {et~al.}(2023){Franciolini}, {Iovino}, {Vaskonen},
  \& {Veerm{\"a}e}}]{2023PhRvL.131t1401F}
{Franciolini}, G., {Iovino}, A.~J., {Vaskonen}, V., \& {Veerm{\"a}e}, H. 2023,
  \bibinfo{title}{{Recent Gravitational Wave Observation by Pulsar Timing
  Arrays and Primordial Black Holes: The Importance of Non-Gaussianities},}
  \prl, 131, 201401, \dodoi{10.1103/PhysRevLett.131.201401}

\bibitem[{T. {Fujita} {et~al.}(2019){Fujita}, {Kuroyanagi}, {Mizuno}, \&
  {Mukohyama}}]{2019PhLB..789..215F}
{Fujita}, T., {Kuroyanagi}, S., {Mizuno}, S., \& {Mukohyama}, S. 2019,
  \bibinfo{title}{{Blue-tilted primordial gravitational waves from massive
  gravity},} Physics Letters B, 789, 215,
  \dodoi{10.1016/j.physletb.2018.12.025}

\bibitem[{S. {Gariazzo} {et~al.}(2019){Gariazzo}, {de Salas}, \&
  {Pastor}}]{2019JCAP...07..014G}
{Gariazzo}, S., {de Salas}, P.~F., \& {Pastor}, S. 2019,
  \bibinfo{title}{{Thermalisation of sterile neutrinos in the early universe in
  the 3+1 scheme with full mixing matrix},} \jcap, 2019, 014,
  \dodoi{10.1088/1475-7516/2019/07/014}

\bibitem[{W. {Giar{\`e}} {et~al.}(2023){Giar{\`e}}, {Forconi}, {Di Valentino},
  \& {Melchiorri}}]{2023MNRAS.520.1757G}
{Giar{\`e}}, W., {Forconi}, M., {Di Valentino}, E., \& {Melchiorri}, A. 2023,
  \bibinfo{title}{{Towards a reliable calculation of relic radiation from
  primordial gravitational waves},} \mnras, 520, 1757,
  \dodoi{10.1093/mnras/stad258}

\bibitem[{M. {Giovannini}(1998){Giovannini}}]{1998PhRvD..58h3504G}
{Giovannini}, M. 1998, \bibinfo{title}{{Gravitational wave constraints on
  post-inflationary phases stiffer than radiation},} \prd, 58, 083504,
  \dodoi{10.1103/PhysRevD.58.083504}

\bibitem[{M. {Giovannini}(2008){Giovannini}}]{2008PhLB..668...44G}
{Giovannini}, M. 2008, \bibinfo{title}{{Stochastic backgrounds of relic
  gravitons, T{\ensuremath{\Lambda}}CDM paradigm and the stiff ages},} Physics
  Letters B, 668, 44, \dodoi{10.1016/j.physletb.2008.07.107}

\bibitem[{Y. {Gong} {et~al.}(2021){Gong}, {Luo}, \&
  {Wang}}]{2021NatAs...5..881G}
{Gong}, Y., {Luo}, J., \& {Wang}, B. 2021, \bibinfo{title}{{Concepts and status
  of Chinese space gravitational wave detection projects},} Nature Astronomy,
  5, 881, \dodoi{10.1038/s41550-021-01480-3}

\bibitem[{Y. {Gouttenoire} {et~al.}(2021){Gouttenoire}, {Servant}, \&
  {Simakachorn}}]{2021arXiv211101150G}
{Gouttenoire}, Y., {Servant}, G., \& {Simakachorn}, P. 2021,
  \bibinfo{title}{{Kination cosmology from scalar fields and gravitational-wave
  signatures},} arXiv e-prints, arXiv:2111.01150,
  \dodoi{10.48550/arXiv.2111.01150}

\bibitem[{L.~P. {Grishchuk}(1974){Grishchuk}}]{1974ZhETF..67..825G}
{Grishchuk}, L.~P. 1974, \bibinfo{title}{{Amplification of gravitational waves
  in an isotropic universe},} Zhurnal Eksperimentalnoi i Teoreticheskoi Fiziki,
  67, 825

\bibitem[{W.~J. {Handley} {et~al.}(2015{\natexlab{a}}){Handley}, {Hobson}, \&
  {Lasenby}}]{2015MNRAS.450L..61H}
{Handley}, W.~J., {Hobson}, M.~P., \& {Lasenby}, A.~N. 2015{\natexlab{a}},
  \bibinfo{title}{{polychord: nested sampling for cosmology},} \mnras, 450,
  L61, \dodoi{10.1093/mnrasl/slv047}

\bibitem[{W.~J. {Handley} {et~al.}(2015{\natexlab{b}}){Handley}, {Hobson}, \&
  {Lasenby}}]{2015MNRAS.453.4384H}
{Handley}, W.~J., {Hobson}, M.~P., \& {Lasenby}, A.~N. 2015{\natexlab{b}},
  \bibinfo{title}{{POLYCHORD: next-generation nested sampling},} \mnras, 453,
  4384, \dodoi{10.1093/mnras/stv1911}

\bibitem[{R.~W. {Hellings} \& G.~S. {Downs}(1983){Hellings} \&
  {Downs}}]{1983ApJ...265L..39H}
{Hellings}, R.~W., \& {Downs}, G.~S. 1983, \bibinfo{title}{{Upper limits on the
  isotropic gravitational radiation background from pulsar timing analysis.},}
  \apjl, 265, L39, \dodoi{10.1086/183954}

\bibitem[{M. {Hindmarsh} {et~al.}(2014){Hindmarsh}, {Huber}, {Rummukainen}, \&
  {Weir}}]{2014PhRvL.112d1301H}
{Hindmarsh}, M., {Huber}, S.~J., {Rummukainen}, K., \& {Weir}, D.~J. 2014,
  \bibinfo{title}{{Gravitational Waves from the Sound of a First Order Phase
  Transition},} \prl, 112, 041301, \dodoi{10.1103/PhysRevLett.112.041301}

\bibitem[{A.~H. {Jaffe} \& D.~C. {Backer}(2003){Jaffe} \&
  {Backer}}]{2003ApJ...583..616J}
{Jaffe}, A.~H., \& {Backer}, D.~C. 2003, \bibinfo{title}{{Gravitational Waves
  Probe the Coalescence Rate of Massive Black Hole Binaries},} \apj, 583, 616,
  \dodoi{10.1086/345443}

\bibitem[{K. {Jedamzik} {et~al.}(2010){Jedamzik}, {Lemoine}, \&
  {Martin}}]{2010JCAP...04..021J}
{Jedamzik}, K., {Lemoine}, M., \& {Martin}, J. 2010,
  \bibinfo{title}{{Generation of gravitational waves during early structure
  formation between cosmic inflation and reheating},} \jcap, 2010, 021,
  \dodoi{10.1088/1475-7516/2010/04/021}

\bibitem[{H. {Jeffreys}(1961){Jeffreys}}]{1961Jeffreys}
{Jeffreys}, H. 1961, {Theory of Probability} (Oxford University Press)

\bibitem[{M. {Joyce}(1997){Joyce}}]{1997PhRvD..55.1875J}
{Joyce}, M. 1997, \bibinfo{title}{{Electroweak baryogenesis and the expansion
  rate of the Universe},} \prd, 55, 1875, \dodoi{10.1103/PhysRevD.55.1875}

\bibitem[{M. {Kamionkowski} {et~al.}(1994){Kamionkowski}, {Kosowsky}, \&
  {Turner}}]{1994PhRvD..49.2837K}
{Kamionkowski}, M., {Kosowsky}, A., \& {Turner}, M.~S. 1994,
  \bibinfo{title}{{Gravitational radiation from first-order phase
  transitions},} \prd, 49, 2837, \dodoi{10.1103/PhysRevD.49.2837}

\bibitem[{S. {Khlebnikov} \& I. {Tkachev}(1997){Khlebnikov} \&
  {Tkachev}}]{1997PhRvD..56..653K}
{Khlebnikov}, S., \& {Tkachev}, I. 1997, \bibinfo{title}{{Relic gravitational
  waves produced after preheating},} \prd, 56, 653,
  \dodoi{10.1103/PhysRevD.56.653}

\bibitem[{T. {Kite} {et~al.}(2022){Kite}, {Chluba}, {Ravenni}, \&
  {Patil}}]{2022MNRAS.509.1366K}
{Kite}, T., {Chluba}, J., {Ravenni}, A., \& {Patil}, S.~P. 2022,
  \bibinfo{title}{{Clarifying transfer function approximations for the
  large-scale gravitational wave background in {\ensuremath{\Lambda}}CDM},}
  \mnras, 509, 1366, \dodoi{10.1093/mnras/stab3125}

\bibitem[{S. {Kuroyanagi} {et~al.}(2011){Kuroyanagi}, {Nakayama}, \&
  {Saito}}]{2011PhRvD..84l3513K}
{Kuroyanagi}, S., {Nakayama}, K., \& {Saito}, S. 2011,
  \bibinfo{title}{{Prospects for determination of thermal history after
  inflation with future gravitational wave detectors},} \prd, 84, 123513,
  \dodoi{10.1103/PhysRevD.84.123513}

\bibitem[{L. {Lancaster} {et~al.}(2017){Lancaster}, {Cyr-Racine}, {Knox}, \&
  {Pan}}]{2017JCAP...07..033L}
{Lancaster}, L., {Cyr-Racine}, F.-Y., {Knox}, L., \& {Pan}, Z. 2017,
  \bibinfo{title}{{A tale of two modes: neutrino free-streaming in the early
  universe},} \jcap, 2017, 033, \dodoi{10.1088/1475-7516/2017/07/033}

\bibitem[{P.~D. {Lasky} {et~al.}(2016){Lasky}, {Mingarelli}, {Smith}, {Giblin},
  {Thrane}, {Reardon}, {Caldwell}, {Bailes}, {Bhat}, {Burke-Spolaor}, {Dai},
  {Dempsey}, {Hobbs}, {Kerr}, {Levin}, {Manchester}, {Os{\l}owski}, {Ravi},
  {Rosado}, {Shannon}, {Spiewak}, {van Straten}, {Toomey}, {Wang}, {Wen},
  {You}, \& {Zhu}}]{2016PhRvX...6a1035L}
{Lasky}, P.~D., {Mingarelli}, C.~M.~F., {Smith}, T.~L., {et~al.} 2016,
  \bibinfo{title}{{Gravitational-Wave Cosmology across 29 Decades in
  Frequency},} Physical Review X, 6, 011035, \dodoi{10.1103/PhysRevX.6.011035}

\bibitem[{B. {Li} {et~al.}(2014){Li}, {Rindler-Daller}, \&
  {Shapiro}}]{2014PhRvD..89h3536L}
{Li}, B., {Rindler-Daller}, T., \& {Shapiro}, P.~R. 2014,
  \bibinfo{title}{{Cosmological constraints on Bose-Einstein-condensed scalar
  field dark matter},} \prd, 89, 083536, \dodoi{10.1103/PhysRevD.89.083536}

\bibitem[{B. {Li} \& P.~R. {Shapiro}(2021){Li} \&
  {Shapiro}}]{2021JCAP...10..024L}
{Li}, B., \& {Shapiro}, P.~R. 2021, \bibinfo{title}{{Precision cosmology and
  the stiff-amplified gravitational-wave background from inflation: NANOGrav,
  Advanced LIGO-Virgo and the Hubble tension},} \jcap, 2021, 024,
  \dodoi{10.1088/1475-7516/2021/10/024}

\bibitem[{B. {Li} {et~al.}(2017){Li}, {Shapiro}, \&
  {Rindler-Daller}}]{2017PhRvD..96f3505L}
{Li}, B., {Shapiro}, P.~R., \& {Rindler-Daller}, T. 2017,
  \bibinfo{title}{{Bose-Einstein-condensed scalar field dark matter and the
  gravitational wave background from inflation: New cosmological constraints
  and its detectability by LIGO},} \prd, 96, 063505,
  \dodoi{10.1103/PhysRevD.96.063505}

\bibitem[{L. {Liu} {et~al.}(2023){Liu}, {Chen}, \&
  {Huang}}]{2023JCAP...11..071L}
{Liu}, L., {Chen}, Z.-C., \& {Huang}, Q.-G. 2023, \bibinfo{title}{{Probing the
  equation of state of the early Universe with pulsar timing arrays},} \jcap,
  2023, 071, \dodoi{10.1088/1475-7516/2023/11/071}

\bibitem[{L. {Liu} {et~al.}(2024){Liu}, {Chen}, \&
  {Huang}}]{2024PhRvD.109f1301L}
{Liu}, L., {Chen}, Z.-C., \& {Huang}, Q.-G. 2024, \bibinfo{title}{{Implications
  for the non-Gaussianity of curvature perturbation from pulsar timing
  arrays},} \prd, 109, L061301, \dodoi{10.1103/PhysRevD.109.L061301}

\bibitem[{P.~D. {Meerburg} {et~al.}(2015){Meerburg}, {Hlo{\v{z}}ek},
  {Hadzhiyska}, \& {Meyers}}]{2015PhRvD..91j3505M}
{Meerburg}, P.~D., {Hlo{\v{z}}ek}, R., {Hadzhiyska}, B., \& {Meyers}, J. 2015,
  \bibinfo{title}{{Multiwavelength constraints on the inflationary consistency
  relation},} \prd, 91, 103505, \dodoi{10.1103/PhysRevD.91.103505}

\bibitem[{C.~J. {Moore} \& A. {Vecchio}(2021){Moore} \&
  {Vecchio}}]{2021NatAs...5.1268M}
{Moore}, C.~J., \& {Vecchio}, A. 2021, \bibinfo{title}{{Ultra-low-frequency
  gravitational waves from cosmological and astrophysical processes},} Nature
  Astronomy, 5, 1268, \dodoi{10.1038/s41550-021-01489-8}

\bibitem[{R. {Namba} {et~al.}(2016){Namba}, {Peloso}, {Shiraishi}, {Sorbo}, \&
  {Unal}}]{2016JCAP...01..041N}
{Namba}, R., {Peloso}, M., {Shiraishi}, M., {Sorbo}, L., \& {Unal}, C. 2016,
  \bibinfo{title}{{Scale-dependent gravitational waves from a rolling axion},}
  \jcap, 2016, 041, \dodoi{10.1088/1475-7516/2016/01/041}

\bibitem[{S. {Pi} {et~al.}(2024){Pi}, {Sasaki}, {Wang}, \&
  {Wang}}]{2024PhRvD.110j3529P}
{Pi}, S., {Sasaki}, M., {Wang}, A., \& {Wang}, J. 2024,
  \bibinfo{title}{{Revisiting the ultraviolet tail of the primordial
  gravitational wave},} \prd, 110, 103529, \dodoi{10.1103/PhysRevD.110.103529}

\bibitem[{ {Planck Collaboration} {et~al.}(2020{\natexlab{a}}){Planck
  Collaboration}, {Akrami}, {Arroja}, {Ashdown}, {Aumont}, {Baccigalupi},
  {Ballardini}, {Banday}, {Barreiro}, {Bartolo}, {Basak}, {Benabed}, {Bernard},
  {Bersanelli}, {Bielewicz}, {Bock}, {Bond}, {Borrill}, {Bouchet}, {Boulanger},
  {Bucher}, {Burigana}, {Butler}, {Calabrese}, {Cardoso}, {Carron},
  {Challinor}, {Chiang}, {Colombo}, {Combet}, {Contreras}, {Crill}, {Cuttaia},
  {de Bernardis}, {de Zotti}, {Delabrouille}, {Delouis}, {Di Valentino},
  {Diego}, {Donzelli}, {Dor{\'e}}, {Douspis}, {Ducout}, {Dupac}, {Dusini},
  {Efstathiou}, {Elsner}, {En{\ss}lin}, {Eriksen}, {Fantaye}, {Fergusson},
  {Fernandez-Cobos}, {Finelli}, {Forastieri}, {Frailis}, {Franceschi},
  {Frolov}, {Galeotta}, {Galli}, {Ganga}, {Gauthier}, {G{\'e}nova-Santos},
  {Gerbino}, {Ghosh}, {Gonz{\'a}lez-Nuevo}, {G{\'o}rski}, {Gratton},
  {Gruppuso}, {Gudmundsson}, {Hamann}, {Handley}, {Hansen}, {Herranz}, {Hivon},
  {Hooper}, {Huang}, {Jaffe}, {Jones}, {Keih{\"a}nen}, {Keskitalo}, {Kiiveri},
  {Kim}, {Kisner}, {Krachmalnicoff}, {Kunz}, {Kurki-Suonio}, {Lagache},
  {Lamarre}, {Lasenby}, {Lattanzi}, {Lawrence}, {Le Jeune}, {Lesgourgues},
  {Levrier}, {Lewis}, {Liguori}, {Lilje}, {Lindholm}, {L{\'o}pez-Caniego},
  {Lubin}, {Ma}, {Mac{\'\i}as-P{\'e}rez}, {Maggio}, {Maino}, {Mandolesi},
  {Mangilli}, {Marcos-Caballero}, {Maris}, {Martin},
  {Mart{\'\i}nez-Gonz{\'a}lez}, {Matarrese}, {Mauri}, {McEwen}, {Meerburg},
  {Meinhold}, {Melchiorri}, {Mennella}, {Migliaccio}, {Mitra},
  {Miville-Desch{\^e}nes}, {Molinari}, {Moneti}, {Montier}, {Morgante}, {Moss},
  {M{\"u}nchmeyer}, {Natoli}, {N{\o}rgaard-Nielsen}, {Pagano}, {Paoletti},
  {Partridge}, {Patanchon}, {Peiris}, {Perrotta}, {Pettorino}, {Piacentini},
  {Polastri}, {Polenta}, {Puget}, {Rachen}, {Reinecke}, {Remazeilles}, {Renzi},
  {Rocha}, {Rosset}, {Roudier}, {Rubi{\~n}o-Mart{\'\i}n}, {Ruiz-Granados},
  {Salvati}, {Sandri}, {Savelainen}, {Scott}, {Shellard}, {Shiraishi},
  {Sirignano}, {Sirri}, {Spencer}, {Sunyaev}, {Suur-Uski}, {Tauber},
  {Tavagnacco}, {Tenti}, {Toffolatti}, {Tomasi}, {Trombetti}, {Valiviita}, {Van
  Tent}, {Vielva}, {Villa}, {Vittorio}, {Wandelt}, {Wehus}, {White}, {Zacchei},
  {Zibin}, \& {Zonca}}]{2020A&A...641A..10P}
{Planck Collaboration}, {Akrami}, Y., {Arroja}, F., {et~al.}
  2020{\natexlab{a}}, \bibinfo{title}{{Planck 2018 results. X. Constraints on
  inflation},} \aap, 641, A10, \dodoi{10.1051/0004-6361/201833887}

\bibitem[{ {Planck Collaboration} {et~al.}(2020{\natexlab{b}}){Planck
  Collaboration}, {Aghanim}, {Akrami}, {Ashdown}, {Aumont}, {Baccigalupi},
  {Ballardini}, {Banday}, {Barreiro}, {Bartolo}, {Basak}, {Battye}, {Benabed},
  {Bernard}, {Bersanelli}, {Bielewicz}, {Bock}, {Bond}, {Borrill}, {Bouchet},
  {Boulanger}, {Bucher}, {Burigana}, {Butler}, {Calabrese}, {Cardoso},
  {Carron}, {Challinor}, {Chiang}, {Chluba}, {Colombo}, {Combet}, {Contreras},
  {Crill}, {Cuttaia}, {de Bernardis}, {de Zotti}, {Delabrouille}, {Delouis},
  {Di Valentino}, {Diego}, {Dor{\'e}}, {Douspis}, {Ducout}, {Dupac}, {Dusini},
  {Efstathiou}, {Elsner}, {En{\ss}lin}, {Eriksen}, {Fantaye}, {Farhang},
  {Fergusson}, {Fernandez-Cobos}, {Finelli}, {Forastieri}, {Frailis},
  {Fraisse}, {Franceschi}, {Frolov}, {Galeotta}, {Galli}, {Ganga},
  {G{\'e}nova-Santos}, {Gerbino}, {Ghosh}, {Gonz{\'a}lez-Nuevo}, {G{\'o}rski},
  {Gratton}, {Gruppuso}, {Gudmundsson}, {Hamann}, {Handley}, {Hansen},
  {Herranz}, {Hildebrandt}, {Hivon}, {Huang}, {Jaffe}, {Jones}, {Karakci},
  {Keih{\"a}nen}, {Keskitalo}, {Kiiveri}, {Kim}, {Kisner}, {Knox},
  {Krachmalnicoff}, {Kunz}, {Kurki-Suonio}, {Lagache}, {Lamarre}, {Lasenby},
  {Lattanzi}, {Lawrence}, {Le Jeune}, {Lemos}, {Lesgourgues}, {Levrier},
  {Lewis}, {Liguori}, {Lilje}, {Lilley}, {Lindholm}, {L{\'o}pez-Caniego},
  {Lubin}, {Ma}, {Mac{\'\i}as-P{\'e}rez}, {Maggio}, {Maino}, {Mandolesi},
  {Mangilli}, {Marcos-Caballero}, {Maris}, {Martin}, {Martinelli},
  {Mart{\'\i}nez-Gonz{\'a}lez}, {Matarrese}, {Mauri}, {McEwen}, {Meinhold},
  {Melchiorri}, {Mennella}, {Migliaccio}, {Millea}, {Mitra},
  {Miville-Desch{\^e}nes}, {Molinari}, {Montier}, {Morgante}, {Moss}, {Natoli},
  {N{\o}rgaard-Nielsen}, {Pagano}, {Paoletti}, {Partridge}, {Patanchon},
  {Peiris}, {Perrotta}, {Pettorino}, {Piacentini}, {Polastri}, {Polenta},
  {Puget}, {Rachen}, {Reinecke}, {Remazeilles}, {Renzi}, {Rocha}, {Rosset},
  {Roudier}, {Rubi{\~n}o-Mart{\'\i}n}, {Ruiz-Granados}, {Salvati}, {Sandri},
  {Savelainen}, {Scott}, {Shellard}, {Sirignano}, {Sirri}, {Spencer},
  {Sunyaev}, {Suur-Uski}, {Tauber}, {Tavagnacco}, {Tenti}, {Toffolatti},
  {Tomasi}, {Trombetti}, {Valenziano}, {Valiviita}, {Van Tent}, {Vibert},
  {Vielva}, {Villa}, {Vittorio}, {Wand elt}, {Wehus}, {White}, {White},
  {Zacchei}, \& {Zonca}}]{2020A&A...641A...6P}
{Planck Collaboration}, {Aghanim}, N., {Akrami}, Y., {et~al.}
  2020{\natexlab{b}}, \bibinfo{title}{{Planck 2018 results. VI. Cosmological
  parameters},} \aap, 641, A6, \dodoi{10.1051/0004-6361/201833910}

\bibitem[{M. {Rajagopal} \& R.~W. {Romani}(1995){Rajagopal} \&
  {Romani}}]{1995ApJ...446..543R}
{Rajagopal}, M., \& {Romani}, R.~W. 1995, \bibinfo{title}{{Ultra--Low-Frequency
  Gravitational Radiation from Massive Black Hole Binaries},} \apj, 446, 543,
  \dodoi{10.1086/175813}

\bibitem[{N. {Ramberg} \& L. {Visinelli}(2021){Ramberg} \&
  {Visinelli}}]{2021PhRvD.103f3031R}
{Ramberg}, N., \& {Visinelli}, L. 2021, \bibinfo{title}{{QCD axion and
  gravitational waves in light of NANOGrav results},} \prd, 103, 063031,
  \dodoi{10.1103/PhysRevD.103.063031}

\bibitem[{D.~J. {Reardon} {et~al.}(2023){Reardon}, {Zic}, {Shannon}, {Hobbs},
  {Bailes}, {Di Marco}, {Kapur}, {Rogers}, {Thrane}, {Askew}, {Bhat},
  {Cameron}, {Cury{\l}o}, {Coles}, {Dai}, {Goncharov}, {Kerr}, {Kulkarni},
  {Levin}, {Lower}, {Manchester}, {Mandow}, {Miles}, {Nathan}, {Os{\l}owski},
  {Russell}, {Spiewak}, {Zhang}, \& {Zhu}}]{2023ApJ...951L...6R}
{Reardon}, D.~J., {Zic}, A., {Shannon}, R.~M., {et~al.} 2023,
  \bibinfo{title}{{Search for an Isotropic Gravitational-wave Background with
  the Parkes Pulsar Timing Array},} \apjl, 951, L6,
  \dodoi{10.3847/2041-8213/acdd02}

\bibitem[{W.-H. {Ruan} {et~al.}(2020){Ruan}, {Liu}, {Guo}, {Wu}, \&
  {Cai}}]{2020NatAs...4..108R}
{Ruan}, W.-H., {Liu}, C., {Guo}, Z.-K., {Wu}, Y.-L., \& {Cai}, R.-G. 2020,
  \bibinfo{title}{{The LISA-Taiji network},} Nature Astronomy, 4, 108,
  \dodoi{10.1038/s41550-019-1008-4}

\bibitem[{K. {Saikawa} \& S. {Shirai}(2020){Saikawa} \&
  {Shirai}}]{2020JCAP...08..011S}
{Saikawa}, K., \& {Shirai}, S. 2020, \bibinfo{title}{{Precise WIMP dark matter
  abundance and Standard Model thermodynamics},} \jcap, 2020, 011,
  \dodoi{10.1088/1475-7516/2020/08/011}

\bibitem[{G. {Sato-Polito} {et~al.}(2023){Sato-Polito}, {Zaldarriaga}, \&
  {Quataert}}]{2023arXiv231206756S}
{Sato-Polito}, G., {Zaldarriaga}, M., \& {Quataert}, E. 2023,
  \bibinfo{title}{{Where are NANOGrav's big black holes?},} arXiv e-prints,
  arXiv:2312.06756, \dodoi{10.48550/arXiv.2312.06756}

\bibitem[{P. {Schwaller}(2015){Schwaller}}]{2015PhRvL.115r1101S}
{Schwaller}, P. 2015, \bibinfo{title}{{Gravitational Waves from a Dark Phase
  Transition},} \prl, 115, 181101, \dodoi{10.1103/PhysRevLett.115.181101}

\bibitem[{A. {Sesana} {et~al.}(2004){Sesana}, {Haardt}, {Madau}, \&
  {Volonteri}}]{2004ApJ...611..623S}
{Sesana}, A., {Haardt}, F., {Madau}, P., \& {Volonteri}, M. 2004,
  \bibinfo{title}{{Low-Frequency Gravitational Radiation from Coalescing
  Massive Black Hole Binaries in Hierarchical Cosmologies},} \apj, 611, 623,
  \dodoi{10.1086/422185}

\bibitem[{J. {Skilling}(2006){Skilling}}]{2006Skilling}
{Skilling}, J. 2006, \bibinfo{title}{{Nested sampling for general Bayesian
  computation},} Bayesian Anal., 1, 833, \dodoi{10.1214/06-BA127}

\bibitem[{B. {Spokoiny}(1993){Spokoiny}}]{1993PhLB..315...40S}
{Spokoiny}, B. 1993, \bibinfo{title}{{Deflationary Universe scenario},} Physics
  Letters B, 315, 40, \dodoi{10.1016/0370-2693(93)90155-B}

\bibitem[{A.~A.
  {Starobinski{\v{i}}}(1979){Starobinski{\v{i}}}}]{1979JETPL..30..682S}
{Starobinski{\v{i}}}, A.~A. 1979, \bibinfo{title}{{Spectrum of relic
  gravitational radiation and the early state of the universe},} Soviet Journal
  of Experimental and Theoretical Physics Letters, 30, 682

\bibitem[{A. {Stewart} \& R. {Brandenberger}(2008){Stewart} \&
  {Brandenberger}}]{2008JCAP...08..012S}
{Stewart}, A., \& {Brandenberger}, R. 2008, \bibinfo{title}{{Observational
  constraints on theories with a blue spectrum of tensor modes},} \jcap, 2008,
  012, \dodoi{10.1088/1475-7516/2008/08/012}

\bibitem[{ {The NANOGrav Collaboration}(2023){The NANOGrav
  Collaboration}}]{2023NG15_KDE}
{The NANOGrav Collaboration}. 2023, \bibinfo{title}{{KDE Representations of the
  Gravitational Wave Background Free Spectra Present in the NANOGrav 15-Year
  Dataset},}, v2 Zenodo, \dodoi{10.5281/zenodo.10344086}

\bibitem[{J. {Torrado} \& A. {Lewis}(2019){Torrado} \&
  {Lewis}}]{2019ascl.soft10019T}
{Torrado}, J., \& {Lewis}, A. 2019, \bibinfo{title}{{Cobaya: Bayesian analysis
  in cosmology},}, Astrophysics Source Code Library, record ascl:1910.019

\bibitem[{J. {Torrado} \& A. {Lewis}(2021){Torrado} \&
  {Lewis}}]{2021JCAP...05..057T}
{Torrado}, J., \& {Lewis}, A. 2021, \bibinfo{title}{{Cobaya: code for Bayesian
  analysis of hierarchical physical models},} \jcap, 2021, 057,
  \dodoi{10.1088/1475-7516/2021/05/057}

\bibitem[{S. {Vagnozzi}(2023){Vagnozzi}}]{2023JHEAp..39...81V}
{Vagnozzi}, S. 2023, \bibinfo{title}{{Inflationary interpretation of the
  stochastic gravitational wave background signal detected by pulsar timing
  array experiments},} Journal of High Energy Astrophysics, 39, 81,
  \dodoi{10.1016/j.jheap.2023.07.001}

\bibitem[{R.~L. {Workman} {et~al.}(2022){Workman}, {Burkert}, {Crede},
  {Klempt}, {Thoma}, {Tiator}, {Agashe}, {Aielli}, {Allanach}, {Amsler},
  {Antonelli}, {Aschenauer}, {Asner}, {Baer}, {Banerjee}, {Barnett}, {Baudis},
  {Bauer}, {Beatty}, {Belousov}, {Beringer}, {Bettini}, {Biebel}, {Black},
  {Blucher}, {Bonventre}, {Bryzgalov}, {Buchmuller}, {Bychkov}, {Cahn},
  {Carena}, {Ceccucci}, {Cerri}, {Chivukula}, {Cowan}, {Cranmer}, {Cremonesi},
  {D'Ambrosio}, {Damour}, {de Florian}, {de Gouv{\^e}a}, {DeGrand}, {de Jong},
  {Demers}, {Dobrescu}, {D'Onofrio}, {Doser}, {Dreiner}, {Eerola}, {Egede},
  {Eidelman}, {El-Khadra}, {Ellis}, {Eno}, {Erler}, {Ezhela}, {Fetscher},
  {Fields}, {Freitas}, {Gallagher}, {Gershtein}, {Gherghetta},
  {Gonzalez-Garcia}, {Goodman}, {Grab}, {Gritsan}, {Grojean}, {Groom},
  {Gr{\"u}newald}, {Gurtu}, {Gutsche}, {Haber}, {Hamel}, {Hanhart},
  {Hashimoto}, {Hayato}, {Hebecker}, {Heinemeyer}, {Hern{\'a}ndez-Rey},
  {Hikasa}, {Hisano}, {H{\"o}cker}, {Holder}, {Hsu}, {Huston}, {Hyodo},
  {Ianni}, {Kado}, {Karliner}, {Katz}, {Kenzie}, {Khoze}, {Klein}, {Krauss},
  {Kreps}, {Kri{\v{z}}an}, {Krusche}, {Kwon}, {Lahav}, {Laiho}, {Lellouch},
  {Lesgourgues}, {Liddle}, {Ligeti}, {Lin}, {Lippmann}, {Liss}, {Littenberg},
  {Louren{\c{c}}o}, {Lugovsky}, {Lugovsky}, {Lusiani}, {Makida}, {Maltoni},
  {Mannel}, {Manohar}, {Marciano}, {Masoni}, {Matthews}, {Mei{\ss}ner},
  {Melzer-Pellmann}, {Mikhasenko}, {Miller}, {Milstead}, {Mitchell},
  {M{\"o}nig}, {Molaro}, {Moortgat}, {Moskovic}, {Nakamura}, {Narain}, {Nason},
  {Navas}, {Nelles}, {Neubert}, {Nevski}, {Nir}, {Olive}, {Patrignani},
  {Peacock}, {Petrov}, {Pianori}, {Pich}, {Piepke}, {Pietropaolo}, {Pomarol},
  {Pordes}, {Profumo}, {Quadt}, {Rabbertz}, {Rademacker}, {Raffelt},
  {Ramsey-Musolf}, {Ratcliff}, {Richardson}, {Ringwald}, {Robinson}, {Roesler},
  {Rolli}, {Romaniouk}, {Rosenberg}, {Rosner}, {Rybka}, {Ryskin}, {Ryutin},
  {Sakai}, {Sarkar}, {Sauli}, {Schneider}, {Sch{\"o}nert}, {Scholberg},
  {Schwartz}, {Schwiening}, {Scott}, {Sefkow}, {Seljak}, {Sharma}, {Sharpe},
  {Shiltsev}, {Signorelli}, {Silari}, {Simon}, {Sj{\"o}strand}, {Skands},
  {Skwarnicki}, {Smoot}, {Soffer}, {Sozzi}, {Spanier}, {Spiering}, {Stahl},
  {Stone}, {Sumino}, {Syphers}, {Takahashi}, {Tanabashi}, {Tanaka},
  {Ta{\v{s}}evsk{\'y}}, {Terao}, {Terashi}, {Terning}, {Thorne}, {Titov},
  {Tkachenko}, {Tovey}, {Trabelsi}, {Urquijo}, {Valencia}, {Van de Water},
  {Varelas}, {Venanzoni}, {Verde}, {Vivarelli}, {Vogel}, {Vogelsang},
  {Vorobyev}, {Wakely}, {Walkowiak}, {Walter}, {Wands}, {Weinberg}, {Weinberg},
  {Wermes}, {White}, {Wiencke}, {Willocq}, {Wohl}, {Woody}, {Yao}, {Yokoyama},
  {Yoshida}, {Zanderighi}, {Zeller}, {Zenin}, {Zhu}, {Zhu}, {Zimmermann},
  {Zyla}, \& {Particle Data Group}}]{2022PTEP.2022h3C01W}
{Workman}, R.~L., {Burkert}, V.~D., {Crede}, V., {et~al.} 2022,
  \bibinfo{title}{{Review of Particle Physics},} Progress of Theoretical and
  Experimental Physics, 2022, 083C01, \dodoi{10.1093/ptep/ptac097}

\bibitem[{H. {Xu} {et~al.}(2023){Xu}, {Chen}, {Guo}, {Jiang}, {Wang}, {Xu},
  {Xue}, {Nicolas Caballero}, {Yuan}, {Xu}, {Wang}, {Hao}, {Luo}, {Lee}, {Han},
  {Jiang}, {Shen}, {Wang}, {Wang}, {Xu}, {Wu}, {Manchester}, {Qian}, {Guan},
  {Huang}, {Sun}, \& {Zhu}}]{2023RAA....23g5024X}
{Xu}, H., {Chen}, S., {Guo}, Y., {et~al.} 2023, \bibinfo{title}{{Searching for
  the Nano-Hertz Stochastic Gravitational Wave Background with the Chinese
  Pulsar Timing Array Data Release I},} Research in Astronomy and Astrophysics,
  23, 075024, \dodoi{10.1088/1674-4527/acdfa5}

\bibitem[{X. {Xue} {et~al.}(2021){Xue}, {Bian}, {Shu}, {Yuan}, {Zhu}, {Bhat},
  {Dai}, {Feng}, {Goncharov}, {Hobbs}, {Howard}, {Manchester}, {Russell},
  {Reardon}, {Shannon}, {Spiewak}, {Thyagarajan}, \&
  {Wang}}]{2021PhRvL.127y1303X}
{Xue}, X., {Bian}, L., {Shu}, J., {et~al.} 2021, \bibinfo{title}{{Constraining
  Cosmological Phase Transitions with the Parkes Pulsar Timing Array},} \prl,
  127, 251303, \dodoi{10.1103/PhysRevLett.127.251303}

\bibitem[{G. {Ye} {et~al.}(2024){Ye}, {Zhu}, \& {Cai}}]{2024JHEP...02..008Y}
{Ye}, G., {Zhu}, M., \& {Cai}, Y. 2024, \bibinfo{title}{{Null energy condition
  violation during inflation and pulsar timing array observations},} Journal of
  High Energy Physics, 2024, 8, \dodoi{10.1007/JHEP02(2024)008}

\bibitem[{F. {Zhang} {et~al.}(2025){Zhang}, {Luo}, {Li}, {Cao}, {Peng},
  {Meyers}, \& {Shapiro}}]{2025arXiv250404054Z}
{Zhang}, F., {Luo}, Y., {Li}, B., {et~al.} 2025, \bibinfo{title}{{SageNet: Fast
  Neural Network Emulation of the Stiff-amplified Gravitational Waves from
  Inflation},} arXiv e-prints, arXiv:2504.04054,
  \dodoi{10.48550/arXiv.2504.04054}

\end{thebibliography}
\bibliographystyle{aasjournalv7}



\end{document}